\documentstyle [12pt,epsf,rotate] {article}

\newcommand{\nwc}{\newcommand}
\nwc{\cl}  {$\clubsuit$}

\nwc{\hyp} {\hyphenation}
\nwc{\be}  {\begin{equation}}
\nwc{\ee}  {\end{equation}}
\nwc{\ba}  {\begin{array}}
\nwc{\ea}  {\end{array}}
\nwc{\bdm} {\begin{displaymath}}
\nwc{\edm} {\end{displaymath}}
\nwc{\bea} {\be\ba{rcl}}
\nwc{\eea} {\ea\ee}
\nwc{\beas}{\begin{eqnarray*}}
\nwc{\eeas}{\end{eqnarray*}}
\nwc{\ben} {\begin{eqnarray}}
\nwc{\een} {\end{eqnarray}}
\nwc{\bda} {\bdm\ba{lcl}}
\nwc{\eda} {\ea\edm}
\nwc{\bc}  {\begin{center}}
\nwc{\ec}  {\end{center}}
\nwc{\ds}  {\displaystyle}
\nwc{\bmat}{\left(\ba}
\nwc{\emat}{\ea\right)}
\nwc{\non} {\nonumber}
\nwc{\bib} {\bibitem}
\nwc{\lra} {\longrightarrow}
\nwc{\Llra}{\Longleftrightarrow}
\nwc{\ra}  {\rightarrow}
\nwc{\Ra}  {\Rightarrow}
\nwc{\lmt} {\longmapsto}
\nwc{\prl} {\partial}
\nwc{\iy}  {\infty}
\nwc{\ol}  {\overline}
\nwc{\hm}  {\hspace{3mm}}
\nwc{\lf}  {\left}
\nwc{\ri}  {\right}
\nwc{\lm}  {\limits}
\nwc{\lb}  {\lbrack}
\nwc{\rb}  {\rbrack}
\nwc{\ov}  {\over}
\nwc{\pri}  {\prime}
\nwc{\nnn} {\nonumber \vspace{.2cm} \\ }
\nwc{\Sc}  {{\cal S}}
\nwc{\Lc}  {{\cal L}}
\nwc{\Rc}  {{\cal R}}
\nwc{\Dc}  {{\cal D}}
\nwc{\Oc}  {{\cal O}}
\nwc{\Cc}  {{\cal C}}
\nwc{\Pc}  {{\cal P}}
\nwc{\Mc}  {{\cal M}}
\nwc{\Ec}  {{\cal E}}
\nwc{\Fc}  {{\cal F}}
\nwc{\Hc}  {{\cal H}}
\nwc{\Kc}  {{\cal K}}
\nwc{\Xc}  {{\cal X}}
\nwc{\Gc}  {{\cal G}}
\nwc{\Zc}  {{\cal Z}}
\nwc{\Nc}  {{\cal N}}
\nwc{\fca} {{\cal f}}
\nwc{\xc}  {{\cal x}}
\nwc{\Ac}  {{\cal A}}
\nwc{\Bc}  {{\cal B}}
\nwc{\Uc}  {{\cal U}}
\nwc{\Vc}  {{\cal V}}
\nwc{\Th} {\Theta}
\nwc{\th} {\theta}
\nwc{\vth} {\vartheta}
\nwc{\eps}{\epsilon}
\nwc{\si} {\sigma}
\nwc{\Gm} {\Gamma}
\nwc{\gm} {\gamma}
\nwc{\bt} {\beta}
\nwc{\La} {\Lambda}
\nwc{\la} {\lambda}
\nwc{\om} {\omega}
\nwc{\Om} {\Omega}
\nwc{\dt} {\delta}
\nwc{\Si} {\Sigma}
\nwc{\Dt} {\Delta}
\nwc{\al} {\alpha}
\nwc{\vp} {\varphi}
\nwc{\kp} {\kappa}
\nwc{\Id}  {{\bf 1}}
\nwc{\diag} {{\rm diag}}
\nwc{\inv}  {{\rm inv}}
\nwc{\mod}  {{\rm mod}}
\nwc{\hal} {\frac{1}{2}}
\nwc{\tpi}  {2\pi i}
\def \lta {\mathrel{\vcenter
     {\hbox{$<$}\nointerlineskip\hbox{$\sim$}}}}

\newsavebox{\nnin} \sbox{\nnin}{$\hspace{1mm}\in\kern -.8em /
                   \hspace{1mm}$}

\newcommand{\sub}{\subset}
\newsavebox{\nnsub} \sbox{\nnsub}{$\hspace{1mm}\sub\kern -.9em /
            \hspace{1mm}$}

\def\KK{{\rm I\kern -.2em  K}}
\def\NN{{\rm I\kern -.16em N}}
\def\RR{{\rm I\kern -.2em  R}}
\def\ZZ{Z \kern -.43em Z}
\def\QQ{{\rm \kern .25em
             \vrule height1.4ex depth-.12ex width.06em\kern-.31em Q}}
\def\CC{{\rm \kern .25em
             \vrule height1.4ex depth-.12ex width.06em\kern-.31em C}}
\def\ZZZ{Z\kern -0.31em Z}

\nwc{\olnu}  {\ol{\nu}}
\nwc{\olla}  {\ol{\la}}
\nwc{\olm}   {\ol{m}}
\nwc{\olmu}  {\ol{\mu}}
\nwc{\olh}   {\ol{h}}
\nwc{\olpsi} {\ol{\psi}}
\nwc{\olsi}  {\ol{\sigma}}
\nwc{\olgm}  {\ol{\gm}}
\nwc{\prlt}  {\frac{\prl}{\prl t}}
\nwc{\ttau}  {\tilde{\tau}}
\nwc{\trho}  {\tilde{\rho}}
\nwc{\tP}    {\tilde{P}}
\nwc{\tU}    {\tilde{U}}
\nwc{\teps}  {\tilde{\eps}}
\nwc{\tla}   {\tilde{\la}}
\nwc{\tit}    {\tilde{t}}
\nwc{\iddq}  {\int\frac{d^dq}{(2\pi)^d}}
\nwc{\prpr}  {\prime\prime}
\nwc{\rN}    {\left(\frac{\rho}{N}\right)}
\nwc{\rNt}    {\left(\frac{\rho}{N}\right)^{\frac{N-2}{2}}}
\nwc{\rnN}   {\left(\frac{\rho_0}{N}\right)}
\nwc{\rnNt}    {\left(\frac{\rho_0}{N}\right)^{\frac{N-2}{2}}}
\nwc{\rnNf}    {\left(\frac{\rho_0}{N}\right)^{\frac{N-4}{2}}}
\nwc{\rNs}    {\left(\frac{\rho_0}{N}\right)^{\frac{N-6}{2}}}
\nwc{\kNt}    {\left(\frac{\kappa}{N}\right)^{\frac{N-2}{2}}}
\nwc{\kNf}    {\left(\frac{\kappa}{N}\right)^{\frac{N-4}{2}}}
\nwc{\kNs}    {\left(\frac{\kappa}{N}\right)^{\frac{N-6}{2}}}

\newcounter{app}
\def\app{\par
 \addtocounter{app}{1}
 \def\thesection{\Alph{app}}
 \def\ksection{\Alph{app}}}

\def\appendix#1{\app\sect{#1}}

\newcounter{Folgeformel}
\newcommand{\sect}[1]{ \section{#1} \setcounter{equation}{0} }
\begin{document}

\begin{titlepage}

\title{ Equation of state near the endpoint of the critical line }
%\thanks{Supported by }

\author{{\sc S.\ Seide} \\
 \\ and \\ \\
{\sc C.\ Wetterich\thanks{Email: C.Wetterich@thphys.uni-heidelberg.de}}
\\ \\ \\
{\em Institut f\"ur Theoretische Physik} \\
{\em Universit\"at Heidelberg} \\
{\em Philosophenweg 16} \\
{\em 69120 Heidelberg, Germany}}

%\date{}
\maketitle

\begin{picture}(5,2.5)(-350,-450)
\put(12,-105){HD--THEP--98--20}
%\put(12,-120){hep-th/9609019}
\end{picture}

\thispagestyle{empty}

\begin{abstract}
We discuss first order transitions for systems in the Ising universality class.
The critical long distance physics near the endpoint of the critical line is
explicitly connected to microscopic properties of a given system. Information
about the short distance physics can therefore be extracted from the precise
location of the endpoint and non-universal amplitudes. Our method is based on
non-perturbative flow equations and yields directly the universal features of
the equation of state, without additional theoretical assumptions of scaling
or resummations of perturbative series. The universal results compare well
with other methods.

\end{abstract}

\end{titlepage}

% *****************************************************************************
\sect{Introduction \label{intro}}

Many phase transitions are described near the critical
 temperature by a one-component scalar field theory without internal
symmetries. A typical example is the water-vapour transition where the
field $\vp(x)$ corresponds to the average density field $n(x)$.
At normal pressure one
observes a first order transition corresponding to a jump in $\vp$ from
high (water) to low (vapour) values as the temperature $T$ is increased.
With increasing pressure the first order transition line ends at some
critical pressure $p_{*}$ in an endpoint. For $p > p_{*}$ the phase
transition is replaced by an analytical crossover.

This behaviour is common to many systems and characterizes the universality
class of the Ising model. As another example from particle physics, the high
temperature electroweak phase transition in the early universe is described
by this universality class if the mass of the Higgs particle in the standard
model is near the endpoint value \mbox{$M_{H*}\approx 72$GeV}\cite{Rummukainen98}.
An Ising type endpoint should also exist if the high temperature
or high density chiral
phase transition in QCD
% \cite{QCDtransit}
or the gas-liquid transition for
nuclear matter are of first order in some region of parameter space.

Very often the location of the endpoint - e.g. the critical $T_{*}$, $p_{*}$
and $n_{*}$ for the liquid-gas transition - is measured quite precisely.
The approach to criticality is governed by universal scaling laws with
critical exponents. Experimental information is also available about the
non-universal amplitudes appearing in this scaling behaviour. These
non-universal critical properties are specific for a given system, and the
question arises how they can be used to gain precise information about the
underlying microscopic physics.
This problem clearly involves the difficult task of an explicit connection
between the short distance physics and the collective behaviour leading to a
very large correlation length.

So far renormalization group methods \cite{KadanoffWilson,otherRGEs,Zinn93}
have established the structure of this relation and led to a precise
determination of the universal critical properties. A suitable implementation
of these ideas should allow us to complete the task by mapping details of
microscopic physics to non-universal critical quantities. In this paper we
show that this is indeed possible. A demonstration is given for the liquid-gas
transition in carbon dioxide.

A very useful concept for the study of phase transitions is the coarse grained
free energy or effective
average action $\Gamma_k[\vp]$ \cite{Wet93,Wet91_93}.
It is related to the (grand canonical)
partition function
in presence of an effective infrared cutoff $R_k$,
\bea
  \ds Z_k[j] & = & \ds
               \int {\mathcal D}\chi \exp\left\{-S[\chi]+\int d^3x
         \left(j(x)\chi(x) -\hal \chi(x)R_k(-\partial^2)\chi(x)\right)\right\},
  \label{erzFktl}
\eea
by a Legendre transform:
\bea
  \ds \Gamma_k[\vp] & = & \ds
      -\ln Z_k[j] + \int d^3x \left( j(x)\vp(x)
                   -\hal\vp(x) R_k(-\partial^2)\vp(x)\right)   \nnn
  \ds \vp(x) & = & \ds \frac{\delta}{\delta j(x)} \ln Z_k[j].
\eea
For the liquid-gas transition
$\chi(x)$ is the microscopic density field, $S[\chi]=H/T$ the microscopic
or classical action and $j(x)$ the source, which will be specified below.
The momentum dependent infrared cutoff obeys $R_k(q^2)\sim k^2$ for
$q^2\ll k^2$ and should vanish rapidly for $q^2\gg k^2$. This assures
that only fluctuations with momenta $q^2 > k^2$ are
effectively included in the functional integral (\ref{erzFktl}).
Below, we will often not consider the most general form of $\Gamma_k$ but
rather work with a truncation which only includes the most general terms
containing up to two derivatives,
\bea
   \Gamma_k[\vp] & = & \ds \int d^3 x \left\{
      U_k(\vp(x)) + \hal Z_k(\vp(x))\partial^{\mu}\vp \partial_{\mu}\vp
   \right\}.
   \label{ansatzGamma}
\eea
Our final aim will be the computation of the potential
$U\equiv U_{k\to 0}$
and the wave function renormalization $Z\equiv Z_{k\to 0}$ for a vanishing
(or very small) infrared cutoff.
In absence of the infrared cutoff the shape of $U(\vp)$ directly determines the
equation of state.
For a homogeneous situation $UT$ corresponds to the free energy density.
Indeed, expressing $U$ as a function of the density one
finds for the liquid-gas system at a given chemical potential $\mu$
\bea
  \ds \frac{\partial U}{\partial n} & = &
  \ds \frac{\mu}{T}.
\eea
Equivalently, one may also use the more familiar form of the equation of state
in terms of the pressure $p$,
\bea
  \ds n^2\frac{\partial}{\partial n}\left(\frac{U}{n}\right) & = &
  \ds \frac{p}{T}.
\eea
(Here the additive constant in $U$ is fixed such that $U(n\!=\!0)\!=\!0$).
The wave function renormalization $Z(\vp)$ contains the
additional information needed for the two point correlation
function at large distance for arbitrary pressure.

A direct computation of $U(\vp)$ and $Z(\vp)$ becomes, however, a difficult
problem for all situations where the correlation length becomes large and
collective phenomena are therefore important. For the water-vapour transition
this happens in the vicinity of the endpoint, i.e. for a pressure and
temperature near $p_{*}$ and $T_{*}$. In this case the free energy can
best be calculated by a stepwise procedure where $k$ is lowered
consecutively. One starts at some large $k$ where only short distance
fluctuations are included and a relatively simple
microphysical computation is reliable. As $k$ is lowered, the effects of
fluctuations with longer wavelengths are added. We will describe the
dependence of the average action on $k$ by a non-perturbative flow equation
which can be derived from an exact renormalization group equation.

In this way the computation of thermodynamic potentials, correlation length
etc. is done in two steps: The first is the computation of a short distance
free energy $\Gamma_{\Lambda}$. This does not involve large length scales
and can be done by a variety of expansion methods or numerical simulations.
This step is not the main emphasis of the present paper and we will use a
relatively crude approximation for the gas-liquid transition.
The second step is more difficult and will be addressed here. It involves
the relation between $\Gamma_{\Lambda}$ and $\Gamma_0$, and has to account
for possible complicated collective long distance fluctuations.

For a large infrared cutoff $k=\Lambda$ one may compute $\Gamma_{\Lambda}$
perturbatively. For example, the lowest order in a virial
expansion for the liquid-gas system yields
\bea
  \lefteqn{\ds U_{\Lambda}(n)  =
   -n\left(1+\ln g + \frac{3}{2}\ln\frac{MT}{2\pi\Lambda^2}\right) } & & \nnn
      &  & + \ds n \ln\left(\frac{n}{(1-b_0(\Lambda)n)\Lambda^3}\right)
      - \frac{b_1(\Lambda)}{T}n^2 + c_{\Lambda}.
  \label{virialU}
\eea
Here $\Lambda^{-1}$ should be of the order of a typical range of intermolecular
interactions, $M$ and $g$ are the mass and the number of degrees of freedom
of a molecule and $b_0$, $b_1$ parameterize the virial coefficient
$\ds B_2(T)=b_0-b_1/T$.
\footnote{
   The Van der Waals coefficients $b_0$, $b_1$ for real gases can be
   found in the literature. These values are valid for small densities.
   They also correspond to
   $k=0$ rather than to
   $k=\Lambda$. Fluctuation effects lead to slightly different values for
   $b_i(\Lambda)$ and $b_i(k=0)$  even away from the critical line.
   We find that these differences are small for $n\ll n_*$.
   Similarly, a constant $c_{\Lambda}$ should be added to $U_{\Lambda}$
   such that $U_0(0)\!=\!0$. }
(The (mass) density $\rho$ is related to the particle density
$n$ by $\rho=Mn$.)
We emphasize that the convergence of a virial expansion is expected to improve
considerably in presence of an infrared cutoff $\Lambda$ which suppresses the
long distance fluctuations.

The field $\vp(x)$ is related to the (space-dependent) particle density
$n(x)$ by
\bea
  \ds \vp(x) & = & K_{\Lambda}(n(x)-\hat{n})
\eea
with $\hat{n}$ some suitable fixed reference density. We approximate the
wave function renormalization $Z_{\Lambda}$ by a constant. It can be inferred
from the correlation length $\hat{\xi}$, evaluated
at some reference density $\hat{n}$ and temperature $\hat{T}$ away
from the critical region, through
\bea
  \ds \hat{\xi}^{-2} & = &
  \ds Z_{\Lambda}^{-1}\frac{\partial^2 U}{\partial\vp^2}
         \left.\right|_{\hat{\vp},\hat{T}}.
\eea
For a suitable scaling factor
\bea
  \ds K_{\Lambda} & = &
  \ds \left(\frac{1}{\hat{n}}+\frac{b_0(2-b_0\hat{n})}{(1-b_0\hat{n})^2}
      -\frac{2b_1}{\hat{T}}\right)^{1/2}\hat{\xi}
\eea
one has $Z_{\Lambda}=1$.

We observe that the terms linear in $n$ in eq. (\ref{virialU}) play only a
role for the relation between $n$ and $\mu$.
It is instructive to subtract from $U_{\Lambda}$ the linear piece in $\vp$
and to expand in powers of $\vp$:
\bea
  \ds U_{\Lambda}(\vp) & = &
  \ds  \frac {m_{\Lambda}^2}{2}\vp^2 + \frac{\gamma_{\Lambda}}{6}\vp^3
      + \frac{\lambda_{\Lambda}}{8}\vp^4 + \ldots
  \label{microsU}

\eea
with
\bea
  \ds m_{\Lambda}^2 & = &
  \ds   K_{\Lambda}^{-2}\left(\frac{1}{\hat{n}}
        +\frac{b_0(2-b_0\hat{n})}{(1-b_0\hat{n})^2}-\frac{2b_1}{T}\right) \nnn
  \ds \gamma_{\Lambda} & = &
  \ds  K_{\Lambda}^{-3}\left(\frac{b_0^2(3-b_0\hat{n})}{(1-b_0\hat{n})^3}
       - \frac{1}{\hat{n}^2}\right)         \nnn
  \ds \lambda_{\Lambda} & = &
  \ds   \frac{2}{3}K_{\Lambda}^{-4}\left(
        \frac{b_0^3(4-b_0\hat{n})}{(1-b_0\hat{n})^4}+ \frac{1}{\hat{n}^3}
             \right).
\eea
For a convenient choice $\ds \hat{n}=\frac{1}{3b_0}$,
$\ds \hat{T}=\frac{8}{11}\frac{b_1}{b_0}$
one has $\gamma_{\Lambda}=0$ and
\bea
  \ds
   K_{\Lambda}=2b_0^{1/2}\hat{\xi},\;\;
   m_{\Lambda}^2=\left(\frac{27}{16}-\frac{b_1}{2b_0T}\right)
   \hat{\xi}^{-2}, \;\;
   \lambda_{\Lambda}=\frac{243}{128}b_0\hat{\xi}^{-4}.
\eea
Typical values for carbon dioxide at the endpoint are
$m_{\Lambda}^2/\Lambda^2=-0.31$,
$\lambda_{\Lambda}/\Lambda=6.63$ for $\Lambda^{-1}=5\cdot 10^{-10}\:m$,
$\hat{\xi}=0.6\:\Lambda^{-1}$.
In the limit (\ref{microsU}) one obtains a $\vp^4$-model. Our explicit
calculations for carbon dioxide will be performed, however, for the
microscopic free energy (\ref{virialU}).
The linear piece in the potential can be absorbed in the source term such that
the equation of state reads
\footnote{
   Note that the source term is independent of $k$. The linear piece in the
   potential can therefore easily be added to $U_{k\to 0}$ once all fluctuation
   effects are included.}
\bea
  \ds \frac{\partial U}{\partial\vp}=j &, &
  \ds j=K_{\Lambda}^{-1}\left(\frac{\mu}{T}+1+\ln g +\frac{3}{2}\ln
      \frac{MT}{2\pi\Lambda^2}\right).
\eea
We emphasize that a polynomial microscopic potential (\ref{microsU}) with
equation
of state $\partial U/\partial\vp = j$ is a good approximation for a large
variety of
different systems. For the example of magnets $\vp$ corresponds to the
magnetization and $jT$ to the external magnetic field. For
$\gamma_{\Lambda}=0$ and $\lambda_{\Lambda}\to\infty$, with finite negative
$m_{\Lambda}^2/\lambda_{\Lambda}$, this is the $Z_2$-symmetric Ising model.

For values of $\vp$ for which the mass term
\mbox{$m^2(\vp)=\frac{1}{Z}\frac{\partial^2 U}{\partial\vp^2}$}
is much larger than $\Lambda^2$ the microscopic approximation to $\Gamma_k$
remains approximately valid also for $k\to 0$, i.e.
$U(\vp)\approx U_{\Lambda}(\vp)$.
The contribution of the long wavelength fluctuations is suppressed by
the small correlation length or large mass.
In the range where \mbox{$m^2(\vp) \ll \Lambda^2$},
however, long distance fluctuations become important and perturbation theory
looses its validity.
Our aim is a description of this region by means of non-perturbative
renormalization group techniques. Beyond the computation of universal
critical exponents and amplitude ratios we want to establish an explicit
connection between the universal critical equation of state and the microscopic
free energy $\Gamma_{\Lambda}$.

\begin{figure}[h]
\unitlength1.0cm
\begin{center}
\begin{picture}(13.,9.)
\put(6.5,-0.5){$\rho \:[g \: cm^{-3}]$ }
\put(1.2,7.){$p \:[bar]$}
\put(-0.5,0.){
\epsfysize=13.cm
\epsfxsize=9.cm
\rotate[r]{\epsffile{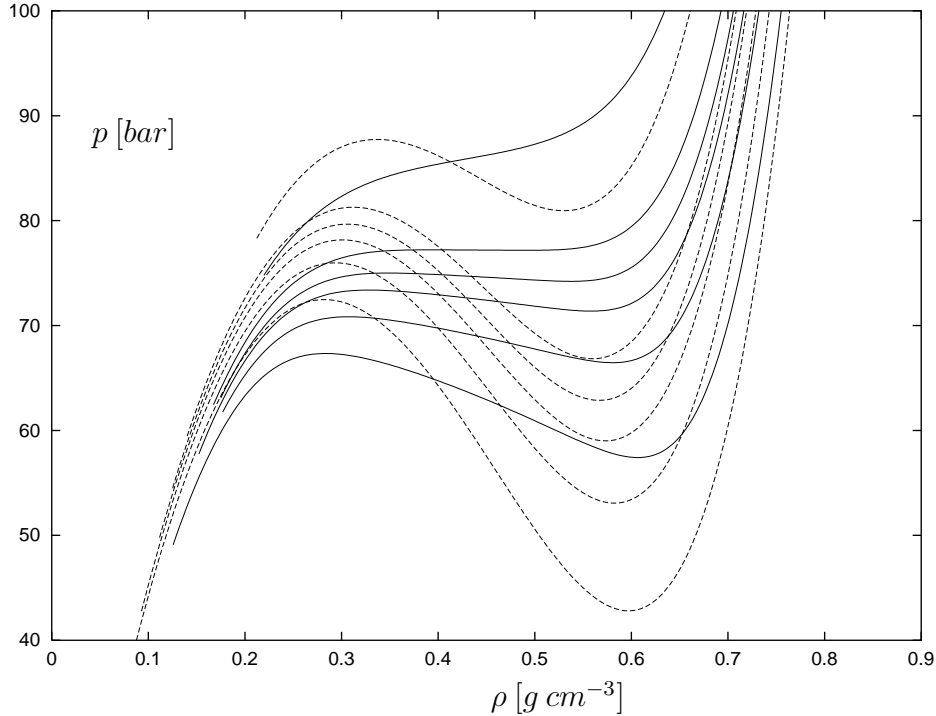}}
}
%\put(0.,0.){\framebox(10.,10.)}
\end{picture}
\end{center}
\caption[]{
\footnotesize $p\!-\!\rho$-isotherms of $CO_2$ for $T=295K$, $T=300K$,
 $T=303K$, $T=305K$, $T=307K$ and $T=315K$. The dashed lines represent the
virial expansion (at the scale
$k\!=\!\Lambda$) in 2nd order with $b_0\!=\!34 \: cm^3\: mol^{-1}$,
$b_1\!=\!3.11\cdot 10^{6} \: bar \: cm^6 \: mol^{-2}$.
The solid lines are the results at the scale $k=0$
 ($\hat{\xi}_{\Lambda}=0.6$).}
\label{CO2plot}
\end{figure}

In fig. \ref{CO2plot} we plot our results for the equation of state near the
endpoint of the critical line for carbon dioxide. For the microscopic scale
we have chosen $\Lambda^{-1}=0.5\: nm$. For $\hat{\xi}=0.6\Lambda^{-1}$,
$b_0(\Lambda)=34\: cm^3 mol^{-1}$,
$b_1(\Lambda)=3.11\cdot 10^{6} \: bar\: cm^6 mol^{-2}$ we find
the location of the endpoint at $T_*=307.4\ K, p_*=77.6\ bar,\ \rho_*
=0.442\ gcm^{-3}$. This compares well with the experimental
values
$T_*=304.15 \; K$, $p_*=73.8 \; bar$, $\rho_*=0.468 \; g cm^{-3}$.
Comparing with literature values $b_i(0)_{ld}$ for low density
this yields $b_0(\Lambda)/b_0(0)_{ld}=0.8$,
$b_1(\Lambda)/b_1(0)_{ld}=0.86$. We conclude that the microscopic
free energy can be approximated reasonably well by a van der Waals form
even for high densities near $n_*$. The coefficients of the virial
expansion are shifted compared to this low density values by 15-20
per cent.
The comparison between the ``microscopic equation of state'' (dashed lines)
and the true equation of state (solid lines) in the plot clearly demonstrates
the importance of the fluctuations in the critical region.
Away from the critical region the fluctuation effects are less significant
and could be computed perturbatively.

%
%
%
%

% Bild:  p-n-Diagramm fuer Kohlendioxid

More details about the equation of state in the critical region as well as
an explicit expression for $U(n)$ will be presented in the following sections.
In sect. \ref{FlowEqus} we briefly review our method and present the relevant
non-perturbative flow equations. Sect. \ref{CritEqofStatIsing} is devoted to
the critical behaviour of the Ising model. We compute the critical equation
of state near the endpoint, critical indices, amplitudes and couplings.
Our results compare well with sophisticated resummation methods of high order
renormalization group improved perturbation theory, the $\epsilon$-expansion
and high temperature series as well as with Monte-Carlo simulations and
experiment. In this section we also give the universal behaviour of the
$\vp$-dependent wave function renormalization. This is needed for the
dependence of the correlation length on density (or a magnetic field in
magnets) and usually not available by other methods.
In sect. \ref{cubicmodel} we turn to the equation of state for first order
transitions. In the present case of the Ising universality class the universal
features can be related to the second order phase transition in the Ising
model.
Such a simple connection is not available for more general systems.

Our flow equation can deal with first order transitions in an arbitrary
context. We have computed here within a $\vp^4$-model with cubic term
and for the microscopic free energy (\ref{virialU}). The perfect agreement
with the Ising model predictions demonstrates the power of this method.
The example of carbon dioxide shows the applicability for
an arbitrary form of the short-distance potential.

%**************************************************************************
\vspace{2cm}
\sect{Flow equation for the effective potential \label{FlowEqus}}

The dependence of the effective average action on the coarse
graining scale $k$ is described by an exact functional differential
evolution equation \cite{Wet93}
\be
 \frac{\partial}{\partial k} \Gm_k [\vp] =
 \hal\int \frac{d^d q}{(2\pi)^d}\left\{\left(
 \Gm_k^{(2)}[\vp]+R_k\right)^{-1}(q^2) \;
 \frac{\prl R_k(q^2)}{\prl k}\right\} .
 \label{ERGE}
\ee
It has the form of a renormalization group improved one-loop equation
in ($d$-dimensional) momentum space.
The diagonal elements of the effective average propagator, i.e.
\[ \left(\Gm_k^{(2)}[\vp]+R_k\right)^{-1}(q^{\prime},q)
 = (\Gm_k^{(2)}+R_k)^{-1}(q^2)(2\pi)^d \delta(q^{\prime}-q)
 + \mathrm{off\!\!-\!\!diagonal\; terms}, \]
involve the matrix of second functional
derivatives of $\Gamma_k$ with respect to the Fourier modes $\vp(q)$,
i.e. $\delta^2\Gm_k[\vp]/\delta\vp(-q^{\prime})\delta\vp(q)
 =\Gm_k^{(2)}[\vp](q^{\prime},q)$.

For a suitable choice of $R_k$ the momentum integral is both infrared and
ultraviolet finite so that no additional regularization is needed and one can
work in arbitrary dimension $d$. This flow equation is formally equivalent to
other versions of exact renormalization group equations
\cite{otherRGEs}. Using instead of the most general form of
$\Gamma$ the truncated ansatz (\ref{ansatzGamma}) one extracts the
(approximate) evolution equation for the potential by evaluating the flow
equation for a constant field $\vp(q)=\varphi(2\pi)^d\delta(q)$,
\bea
   \ds \frac{\partial}{\partial t} U_k(\vp) & = &
   \ds v_d \int_0^{\infty} dx x^{\frac{d}{2}-1}
       \frac{\partial_t R_k(x)}{Z_k(\vp)x+R_k(x)
             +\partial^2 U_k(\vp)/\partial\vp^2}.
   \label{dtU}
\eea
Here \mbox{$t=\ln(k/\Lambda)$}, $x=q^2$,
\mbox{$v_d^{-1} = 2^{d+1} \pi^\frac{d}{2} \Gm\left(d/2\right)$},
and we choose a smooth infrared cutoff
\bea
  \ds R_k(x) & = & \ds \frac{Z_{0,k}\, x}{\exp(x/k^2)-1}.
   \label{StfrmIRCut}
\eea
The flow equation for $Z_k(\vp)$ can be obtained by adding to $\vp$ a small
plane wave with long wavelength (see below). We note that the ansatz
(\ref{ansatzGamma}) can be regarded as the second order in a systematic
derivative expansion for the solution of eq. (\ref{ERGE}).
\cite{propthreshfuncs,Morris94}

Our aim is a numerical solution of the partial differential equation
(\ref{dtU}) with given initial conditions at the scale $k\approx \Lambda$.
For this purpose it is convenient to introduce a dimensionless renormalized
field
\bea
   \ds  \tilde{\vp} & = & \ds k^{\frac{2-d}{2}} Z_{0,k}^{1/2} \vp
\eea
with wave function renormalization $Z_{0,k}=Z_k(\vp_0(k))$ taken at the global
potential minimum $\vp_0(k)$.
We also use
\bea
  \ds u_k(\tilde{\vp}) & = & \ds k^{-d} U_k(\vp)   \nnn
  \ds \tilde{z}_k(\tilde{\vp}) & = & Z_{0,k}^{-1} Z_k(\vp)
\eea
and denote by $u^{\prime}$, $\tilde{z}^{\prime}$ the derivatives with
respect to $\tilde{\vp}$. This yields the scaling form of the flow equation
for $u_k^{\prime}$:
%\end{flushleft}
\bea
   \ds \lefteqn{\frac{\partial}{\partial t} u_k^{\prime}(\tilde{\vp})=
   \ds -\hal (d+2-\eta_0)\cdot u_k^{\prime}(\tilde{\vp})
       + \hal (d-2+\eta_0)\tilde{\vp}\cdot u_k^{\prime\prime}
          (\tilde{\vp})} & &  \nnn
   & - & \ds  2v_d\tilde{z}_k^{\prime}(\tilde{\vp})\cdot
       l_1^{d+2}\left(u_k^{\prime\prime}(\tilde{\vp});\eta_0,
                      \tilde{z}_k(\tilde{\vp}))
           - 2v_d u_k^{\prime\prime\prime}(\tilde{\vp}\right)\cdot
       l_1^d\left(u_k^{\prime\prime}(\tilde{\vp});
                  \eta_0,\tilde{z}_k(\tilde{\vp})\right).
   \label{dtu}

\eea
Similarly, the evolution of $\tilde{z}_k$ is described in the truncation
(\ref{ansatzGamma}) by
\bea
   \ds \lefteqn{\frac{\partial}{\partial t} \tilde{z}_k(\tilde{\vp}) =
    \ds \eta_0\cdot \tilde{z}_k(\tilde{\vp})
        + \hal (d-2+\eta_0)\tilde{\vp}\cdot
          \tilde{z}_k^{\prime}(\tilde{\vp}) } & & \nnn
 & - &  \ds \frac{4}{d} v_d \cdot
   u_k^{\prime\prime\prime}(\tilde{\vp})^2
   \cdot m_{4,0}^d\left(u_k^{\prime\prime}(\tilde{\vp});
                    \eta_0,\tilde{z}_k(\tilde{\vp})\right)  \nnn
 & - &  \ds \frac{8}{d} v_d \cdot u_k^{\prime\prime\prime}(\tilde{\vp})
           \tilde{z}_k^{\prime}(\tilde{\vp})
   \cdot m_{4,0}^{d+2}\left(u_k^{\prime\prime}(\tilde{\vp});\eta_0,
                       \tilde{z}_k(\tilde{\vp})\right) \nnn
 & - &  \ds \frac{4}{d} v_d \cdot \tilde{z}_k^{\prime}(\tilde{\vp})^2
   \cdot m_{4,0}^{d+4}\left(u_k^{\prime\prime}(\tilde{\vp});
                         \eta_0,\tilde{z}_k(\tilde{\vp})\right)
  -   \ds 2 v_d \cdot \tilde{z}_k^{\prime\prime}(\tilde{\vp})\cdot
    l_1^d\left(u_k^{\prime\prime}(\tilde{\vp});
             \eta_0,\tilde{z}_k(\tilde{\vp})\right)  \nnn
 & + & \ds 4 v_d \cdot \tilde{z}_k^{\prime}(\tilde{\vp})
           u_k^{\prime\prime\prime}(\tilde{\vp})\cdot
   l_2^d\left(u_k^{\prime\prime}(\tilde{\vp});
            \eta_0,\tilde{z}_k(\tilde{\vp})\right)  \nnn
 &  + & \ds \frac{2}{d}(1+2d) v_d \cdot \tilde{z}_k^{\prime}(\tilde{\vp})^2
   \cdot
   l_2^{d+2}\left(u_k^{\prime\prime}(\tilde{\vp});
               \eta_0,\tilde{z}_k(\tilde{\vp})\right)
 \label{dtz}
\eea
(see ref. \cite{Wet91_93} for the derivation of an analogous equation in a
different language).
Here the mass threshold functions
\bea
  \ds l_{n}^d(u^{\prime\prime};\eta_0,\tilde{z}) & = &
  \ds -\frac{1}{2}k^{2n-d}\cdot
   Z_{0,k}^n \cdot \int_{0}^{\infty}dx x^{\frac{d}{2}-1} \tilde{\partial}_t
   \left\{
      \frac{1}{(P(x)+Z_{0,k} k^2 u^{\prime\prime})^n}
   \right\}  \nnn
  \ds m_{n,0}^d(u^{\prime\prime};\eta_0,\tilde{z}) & = &
  \ds -\frac{1}{2}k^{2(n-1)-d}\cdot
   Z_{0,k}^{n-2} \cdot \int_{0}^{\infty}dx x^{\frac{d}{2}}\tilde{\partial}_t
   \left\{
      \frac{\dot{P}^2(x)}{(P(x)+Z_{0,k} k^2 u^{\prime\prime})^n}
   \right\}
   \label{threshfuncs}
\eea
(with $P(x)= \tilde{z}Z_{0,k} x + R_k(x)$,
 $\dot{P}\equiv\frac{dP}{dx}$ and $\tilde{\partial}_t$ acting only on $R_k$)
describe the suppression of fluctuations once
$u_k^{\prime\prime}(\tilde{\vp})\gg 1$. A discussion of their properties can
be found in \cite{propthreshfuncs,numsolreference}.
The anomalous dimension
\bea
   \ds \eta_{0,k} \equiv -\frac{d}{dt}\ln Z_{0,k} & = & \ds
     -Z_{0,k}^{-1} \frac{\partial}{\partial t} Z_k(\vp_0)
     - Z_{0,k}^{-1}\cdot \frac{\partial Z_k}{\partial\vp}\left.\right|_{\vp_0}
       \cdot \frac{d\vp_0}{dt}
\eea
is determined by the condition $d\tilde{z}(\tilde{\vp}_0)/dt=0$. It appears
linearly in the threshold functions due to $\tilde{\partial}_t$ acting on
$Z_{0,k}$ in the definition (\ref{StfrmIRCut}) of $R_k$.

For a computation of $\eta_0$ we need the evolution of the potential minimum
$\vp_0(k)$, which follows from the condition
\mbox{$\frac{d}{dt}(\partial U_k/\partial\vp(\vp_0(k))=0$},
namely
\bea
   \ds \lefteqn{\frac{d\tilde{\vp}_0}{dt}  =
    \hal (2-d-\eta_{0})\tilde{\vp_0}  } & &  \nnn
    & & \ds +    2v_d\frac{\tilde{z}_k^{\prime}(\tilde{\vp}_0)}
                  {u_k^{\prime\prime}(\tilde{\vp}_0)}\cdot
         l_1^{d+2}(u_k^{\prime\prime}(\tilde{\vp}_0);\eta_0,1)
    +   2v_d\frac{u_k^{\prime\prime\prime}(\tilde{\vp}_0)}
              {u_k^{\prime\prime}(\tilde{\vp}_0)}\cdot
         l_1^d(u_k^{\prime\prime}(\tilde{\vp}_0);\eta_0,1).
   \label{dtphi0}
\eea
One infers an implicit equation for the anomalous dimension $\eta_{0,k}$,
\bea
   \ds \lefteqn{ \eta_0 =  \frac{4}{d} v_d \cdot
   u_k^{\prime\prime\prime}(\tilde{\vp}_0)^2
   \cdot m_{4,0}^d(u_k^{\prime\prime}(\tilde{\vp}_0); \eta_0,1)
   +   \ds \frac{8}{d} v_d \cdot u_k^{\prime\prime\prime}(\tilde{\vp}_0)
   \tilde{z}_k^{\prime}(\tilde{\vp}_0)
   \cdot m_{4,0}^{d+2}(u_k^{\prime\prime}(\tilde{\vp}_0);\eta_0,1) } \nnn
 & & +  \ds \frac{4}{d} v_d \cdot \tilde{z}_k^{\prime}(\tilde{\vp}_0)^2
   \cdot m_{4,0}^{d+4}(u_k^{\prime\prime}(\tilde{\vp}_0);\eta_0,1)
   +  \ds 2 v_d \cdot \tilde{z}_k^{\prime\prime}(\tilde{\vp}_0)\cdot
    l_1^d(u_k^{\prime\prime}(\tilde{\vp}_0);\eta_0,1) \nnn
 & & - \ds 4 v_d \cdot \tilde{z}_k^{\prime}(\tilde{\vp}_0)
       u_k^{\prime\prime\prime}(\tilde{\vp}_0)\cdot
   l_2^d(u_k^{\prime\prime}(\tilde{\vp}_0);\eta_0,1) \nnn
 & &  -  \ds \frac{2}{d}(1+2d) v_d \cdot \tilde{z}_k^{\prime}(\tilde{\vp}_0)^2
   \cdot
   l_2^{d+2}(u_k^{\prime\prime}(\tilde{\vp}_0);\eta_0,1)   \nnn
 & & -  \ds 2v_d \frac{\tilde{z}_k^{\prime}(\tilde{\vp}_0)}
       {u_k^{\prime\prime}(\tilde{\vp}_0)}
       \cdot \left\{\tilde{z}_k^{\prime}(\tilde{\vp}_0)
       l_1^{d+2}(u_k^{\prime\prime}(\tilde{\vp}_0);\eta_0,1)
       + u_k^{\prime\prime\prime}(\tilde{\vp}_0)
      l_1^d(u_k^{\prime\prime}(\tilde{\vp}_0);\eta_0,1)
       \right\},
   \label{eta0Gl}
\eea
that can be solved by separating the threshold functions in $\eta_0$-dependent
and $\eta_0$-inde\-pen\-dent parts (c.f. eq. (\ref{threshfuncs})).
Since $\eta_0$ will turn out to be only a few
percent,
the neglect of contributions from higher derivative terms not contained
in (\ref{ansatzGamma})
induces a substantial \emph{relative} error for $\eta_0$, despite the good
convergence of the derivative expansion.
We believe that the missing higher derivative contributions to
$\eta_0$ constitute the main uncertainty in our results.

For given initial conditions $U_{\Lambda}(\vp)$, $Z_{\Lambda}(\vp)$ the
system of partial differential equations
(\ref{dtu}),(\ref{dtz}),(\ref{dtphi0}),(\ref{eta0Gl}) can be solved
numerically. A description of the algorithm used for the present work can be
found in \cite{numsolreference}.

%***************************************************************************
\vspace{2cm}
\sect{Critical equation of state for the Ising model \label{CritEqofStatIsing}}

In order to make the discussion transparent we present here first
results for polynomial initial conditions (\ref{microsU}) with
$\tilde{z}_{\Lambda}(\tilde{\vp})=1$.
The term linear in $\vp$ is considered as a source $j$. The special value
$\gamma_{\Lambda}=0$ realizes the $Z_2$-symmetric Ising model.
We start with the results for the universal critical behaviour for this
case.
For this particular purpose we hold $\lambda_{\Lambda}$ fixed and measure the 
deviation from the critical temperature by
\be
   \delta m_{\Lambda}^2=m_{\Lambda}^2-m_{\Lambda,crit}^2
                       =S (T-T_c).
\ee
For the liquid-gas system one has $S=2b_1/(K_{\Lambda}^2 T_c^2)$.
As predicted by general scaling arguments we find that for
small $|\delta m_{\Lambda}^2|/\Lambda^2$ the susceptibility
 $\chi=\overline{m}^{-1}$ and the inverse correlation length
or renormalized mass $m_R=\xi^{-1}$ in the high temperature phase
\mbox{($\delta m_{\Lambda}^2 > 0$)} obey
\bea
   \ds \overline{m}^2 & = & \ds
   (C^{+})^{-1}\cdot (\delta m_{\Lambda}^2)^{\gamma},
\eea
\bea
   \ds m_R & = & \ds
       (\xi^{+})^{-1}\cdot (\delta m_{\Lambda}^2)^{\nu},
\eea
whereas for $T<T_c$ one has
\bea
   \ds \overline{m}^2 & = & \ds
   (C^{-})^{-1}\cdot (-\delta m_{\Lambda}^2)^{\gamma},
\eea
\bea
   \ds m_R & = & \ds
       (\xi^{-})^{-1}\cdot (-\delta m_{\Lambda}^2)^{\nu}.
\eea
Here the renormalized and unrenormalized squared mass terms are defined by
\be
   \overline{m}^2=\frac{\partial^2 U}{\partial\vp^2}(\vp_0),\;\;
   m_R^2=\frac{\partial^2 U}{\partial\vp_R^2}(\vp_0), \;\;
   \vp_R=Z_0^{1/2}\vp.
\ee
While for $T>T_c$ the minimum of $U=U_{k\to 0}$ is at the symmetric
point $\vp_0=0$, the low temperature phase is characterized by spontaneous
symmetry breaking, with
\bea
   \ds \vp_0 & = & \ds B\cdot (-\delta m_{\Lambda}^2)^{\beta},
\eea
\bea
   \ds \vp_{0R}^2 & = & \ds 2 E \cdot (-\delta m_{\Lambda}^2)^{\nu}.
\eea
The anomalous dimension $\eta$ determines the two point function at the
critical temperature and equals $\eta_{0,k}$ for the scaling solution where
\mbox{$\partial_t u=\partial_t \tilde{z}=0$}. Our results for the critical
exponents are compared with those from other methods in table
\ref{IsingExponenten}.
We observe a very good agreement for $\nu$ whereas the relative error for
$\eta$ is comparatively large as expected. Comparison with the lowest order
of the derivative expansion (f) shows a convincing apparent convergence of
this expansion for $\nu$.
For $\eta$ this convergence is hidden by the fact that in \cite{BergTetWet97}
a different determination of $\eta$ was used. Employing the present definition
would lead in lowest order of the derivative expansion to a value $\eta=0.11$.
As expected, the convergence of the derivative expansion is faster for the
very effective exponential cutoff than for the powerlike cutoff (g) which
would lead to unwanted properties of the momentum integrals in the next
order.

\begin{table} [h] \centering{
\begin{tabular}
{|c|c|c|c|c|c|}
\hline
   &  $\nu$ & $\beta$ & $\gamma$ & $\eta$ \\
\hline
 (a) & $0.6304(13)$ & $0.3258(14)$ & 1.2397(13)   &
  $0.0335(25)$ \\
 (b) & $0.6293(26)$ & $0.3260(20)$ & 1.2360(40)   &
  $0.036(6)$  \\
 (c) & $0.6300(49)$ &              & $1.2400(87)$   &
              \\
 (d) & $0.625(1)$  &           &              &
  $0.025(6)$   \\
 (e) &$0.629(3)$  &           &              &
  $0.027(5)$  \\
\hline
 (f) & $0.643$ & $0.336$        & $1.258$ &
  $0.044$  \\
 (g) & $0.6181$ &              &             &
  $0.054$ \\
 (h) & $0.6307$  & $0.3300$  & $1.2322$ &
  $0.0467$  \\
\hline
\hline
 (i) & $0.625(6)$  & $0.316-0.327$ & $1.23-1.25$ &  \\
\hline
\end{tabular} 
\caption{\footnotesize
Critical exponents of the $(d\!=\!3)$-Ising model, calculated with various
methods.\protect\\
(a) Six loop resummed perturbation series at fixed dimension $d=3$
    \protect\cite{Zinn93,GuidaZinn96}. \protect \\
(b) $\epsilon$-expansion in five loop order \protect\cite{Zinn93,GuidaZinn96},
    (c) high temperature series \protect\cite{Reisz95}
    (see also \protect\cite{ZinnLaiFisher96},
    \protect\cite{ButeraComi96}). \protect \\
(d),(e) Monte Carlo simulations \protect\cite {Tsypin94}-
    \protect\cite{CasHas97}. \protect \\
(f)-(h): ``exact'' renormalization group equations.  \protect \\
(f) effective average action for the $O(N)$-model, $N\to 1$,
    with uniform wave function renormalization
    \protect\cite{BergTetWet97} (see also
    \protect\cite{propthreshfuncs}).   \protect \\
(g) scaling solution of equations analogous to (\protect\ref{dtu}),
    (\protect\ref{dtz}) with
    powerlike cutoff \protect \cite{Morris94}. \protect \\
(h) effective average action for one-component scalar field theory
    with field-dependent wave function renormalization (present work).
    \protect \\
(i) experimental data for the liquid-vapour system
    quoted from \protect \cite{Zinn93}
}
}
\label{IsingExponenten}
\end{table}

In order to establish the quantitative connection between the short distance
parameters $m_{\Lambda}^2$ and $\lambda_{\Lambda}$ and the universal
critical behaviour one needs the amplitudes $C^{\pm}$, $\xi^{\pm}$, etc.
For $\lambda_{\Lambda}/\Lambda=5$ we find \mbox{$C^{+}\!=\!1.033$},
 \mbox{$C^{-}\!=\!0.208$},
 \mbox{$\xi^{+}\!=\!0.981$}, \mbox{$\xi^{-}\!=\!0.484$}, \mbox{$B\!=\!0.608$},
 \mbox{$E\!=\!0.208$}.
Here and in the following all dimensionful quantities are quoted in units
of $\Lambda$.
The amplitude $D$ is given by
\mbox{$\partial U_0/\partial\vp =  D\cdot \vp^{\delta}$} on the critical
iso\-therme and we obtain \mbox{$D\!=\!10.213$}.
In table \ref{IsingUnivAmpTab} we present our results for the universal
amplitude ratios
$C^{+}/C^{-}$, $\xi^{+}/\xi^{-}$, \mbox{$R_{\chi}=C^{+}DB^{\delta-1}$},
 \mbox{$\tilde{R}_{\xi}=(\xi^{+})^{\beta/\nu}D^{1/(\delta+1)}B$}.

\begin{table} [h] \centering{
\begin{tabular}
{|c|c|c|c|c|c|c|c|}
\hline
 & $C^{+}/C^{-}$ & $\xi^{+}/\xi^{-}$ & $R_{\chi}$ & $\tilde{R}_{\xi}$ &
 $\xi^{+}E$ & $\lambda_R/m_R$ &
 $\hat{\lambda}_R/\hat{m}_R$ \\
\hline
\hline
 (a) & $4.79\pm 0.10$ &                & $1.669\pm 0.018$ & & & $7.88$ &\\
\hline
 (b) & $4.72\pm 0.14$ &                & $1.648\pm 0.031$ & & & $9.33$ &\\
\hline
 (c) & $4.95 \pm 0.015$ & $1.96 \pm 0.01$ & 1.75 & & & $7.9 - 8.15$ &\\
\hline
 (d) & $4.75\pm 0.03$ & $1.95\pm 0.02$ &                  & & & $7.76$
     & $5.27$ \\
\hline
\hline
 (f) & 4.29 & 1.86 & 1.61 & 0.865 & 0.168  & $9.69$ & $5.55$ \\
\hline
 (h) & 4.966 & 2.027 & 1.647 & 0.903 & 0.204 & $8.11$ & $4.96$ \\
\hline
\hline
 (i) & $4.8-5.2$ &        & $1.69\pm 0.14$ & & & &\\
\hline

\end{tabular} }

%\hspace*{\fill}
\caption{\footnotesize Universal amplitude ratios and couplings of the
$(d\!=\!3)$-Ising model.
 \protect\\
(a) perturbation theory at fixed dimension $d\!=3\!$
    \protect\cite{GuidaZinn96},
(b) $\epsilon$-expansion \protect\cite{GuidaZinn96}. \protect \\
(c) high temperature series. Amplitude ratios from \protect\cite{Zinn93},
    $\lambda_R/m_R$ from \protect\cite{Reisz95,ZinnLaiFisher96,ButeraComi96}.
    \protect \\
(d) Monte Carlo simulations. Amplitude ratios from \protect\cite{CasHas97},
    $\hat{\lambda}_R/\hat{m}_R$ from \protect\cite{Tsypin96},
    $\lambda_R/m_R$ from \protect\cite{Tsypin94}. \protect \\
(f) effective average action for the $O(N)$-model, $N\to 1$,
    with uniform wave function renormalization
    \protect\cite{BergTetWet97}. \protect \\
(h) present work with field-dependent wave function renormalization.\protect \\
(i) experimental data for the liquid-vapour system
    \protect\cite{PrivHohen91}.
\protect \\
}
\label{IsingUnivAmpTab}
\end{table}

The critical exponents and amplitudes only characterize the behaviour of
$U(\vp)$ in the
limits \mbox{$\vp \to \vp_0$} and \mbox{$\vp \to \infty$}. Our method
allows us to compute $U(\vp)$ for arbitrary $\vp$. As an example, the quartic
coupling $\lambda_R=\frac{1}{3}\frac{\partial^4 U}{\partial\vp_R^4}(0)
=\frac{\partial^2U}{\partial\rho_R^2}(0),\ \hat\lambda_R=
\frac{\partial^2U}{\partial\rho^2_R}(\varphi_{0R}),\
\rho_R=\frac{1}{2}\varphi^2_R$,
becomes in the critical region proportional to $m_R$. Our results for
the universal couplings $\lambda_R/m_R$ in the symmetric and
$\hat{\lambda}_R/\hat{m}_R$ in the ordered phase can also be found in
table \ref{IsingUnivAmpTab}.
Here $m_R=\frac{\partial^2 U}{\partial \vp_R^2}\left.\right|_{\vp_R=0}$ in
the symmetric and
$\hat{m}_R=\frac{\partial^2 U}{\partial \vp_R^2}\left.\right|_{\vp_{0R}}$
in the ordered phase.

\begin{figure}[h]
\unitlength1.0cm
\begin{center}
\begin{picture}(13.,9.)
\put(6.4,-0.2){\footnotesize $\vp$ }
\put(1.75,8.){$U_k(\vp)$}

\put(-0.5,0.){
\epsfysize=13.cm
\epsfxsize=9.cm
\rotate[r]{\epsffile{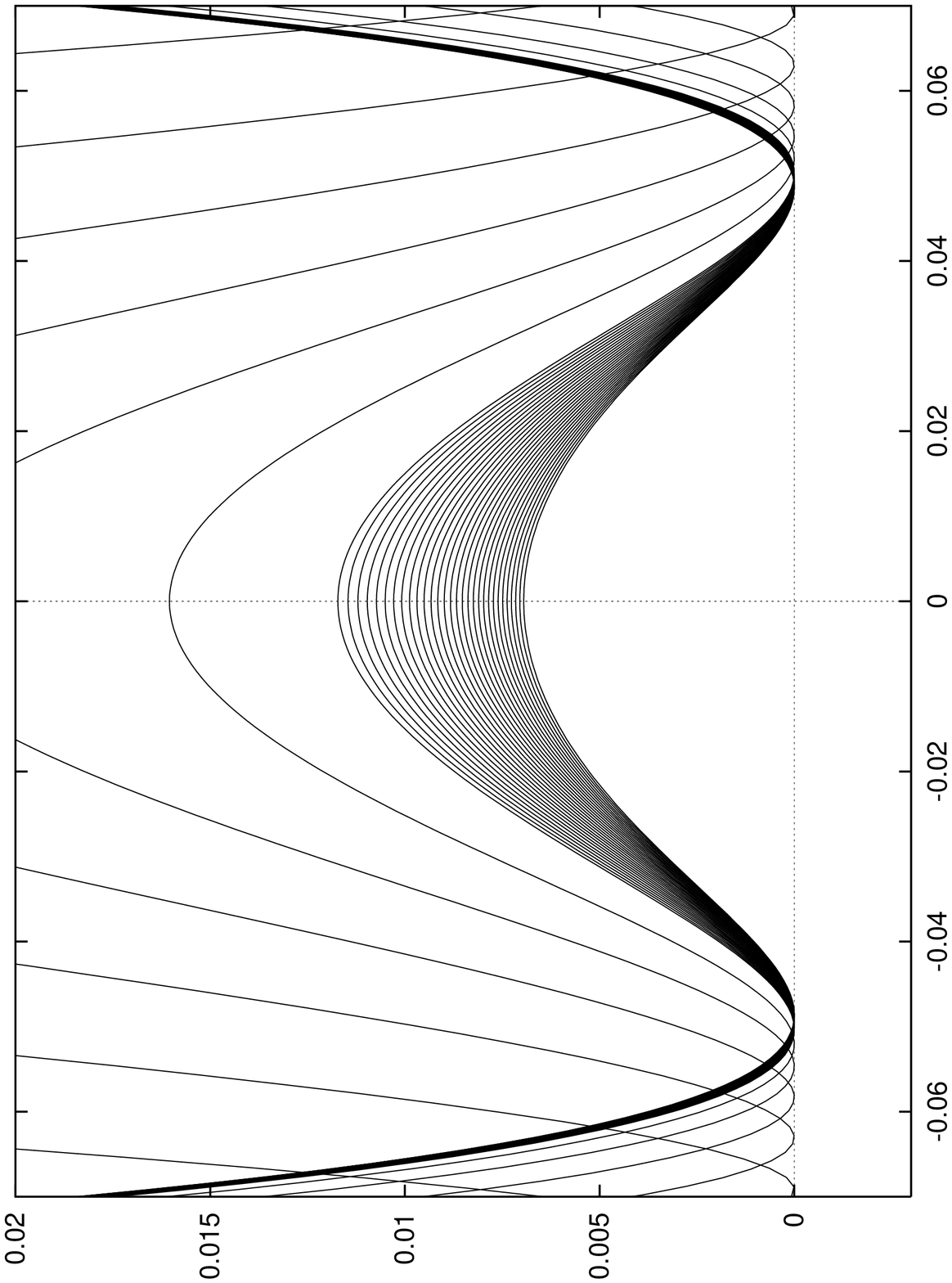}}
}
%\put(0.,0.){\framebox(10.,10.)}
\end{picture}
\end{center}
\caption[]{\label{apptoconvexity}
\footnotesize $U_k(\vp)$ for different scales $k$. The shape of $U$ is stored
 in smaller intervals $\Delta t=-0.02$ after leaving the scaling solution.
 This demonstrates the approach to convexity in the ``inner region'', while
the ``outer region'' becomes $k$-independent.  }
\end{figure}

We should emphasize that the shape of the potential in the low temperature
phase depends on $k$ in the ``inner'' region corresponding to $|\vp|<\vp_0$.
This is due to the fluctuations which are responsible for making the potential
convex in the limit $k\to 0$ \cite{RingWet90,TetWet92,AlexBranchPolon98}.
We illustrate this by plotting the potential for different values of $k$ in
fig. \ref{apptoconvexity}.

% Bild: U in der gebrochenen Phase fuer verschiedene Abbruchskalen

\begin{figure}[h]
\unitlength1.0cm
\begin{center}
\begin{picture}(13.,9.)
%\put(4.,2.){\footnotesize scaling solution}
%\put(4.3,2.){\vector(1,-1){0.4}}
%\put(5.,6.5){\footnotesize cutoff potential $u_{\Lambda}$}
\put(6.,-0.5){$\ds x\cdot f^{-\frac{1}{\beta\delta}}(x)$}
\put(-1.5,9.2){$\ds \frac{\vp}{j^{1/\delta}}=f^{-1/\delta}$}
%\put(7.5,0.8){\footnotesize approach to convexity}
%\put(7.2,1.){\vector(-1,1){0.4}}

\put(-0.5,0.){
\epsfysize=13.cm
\epsfxsize=9.cm
\rotate[r]{\epsffile{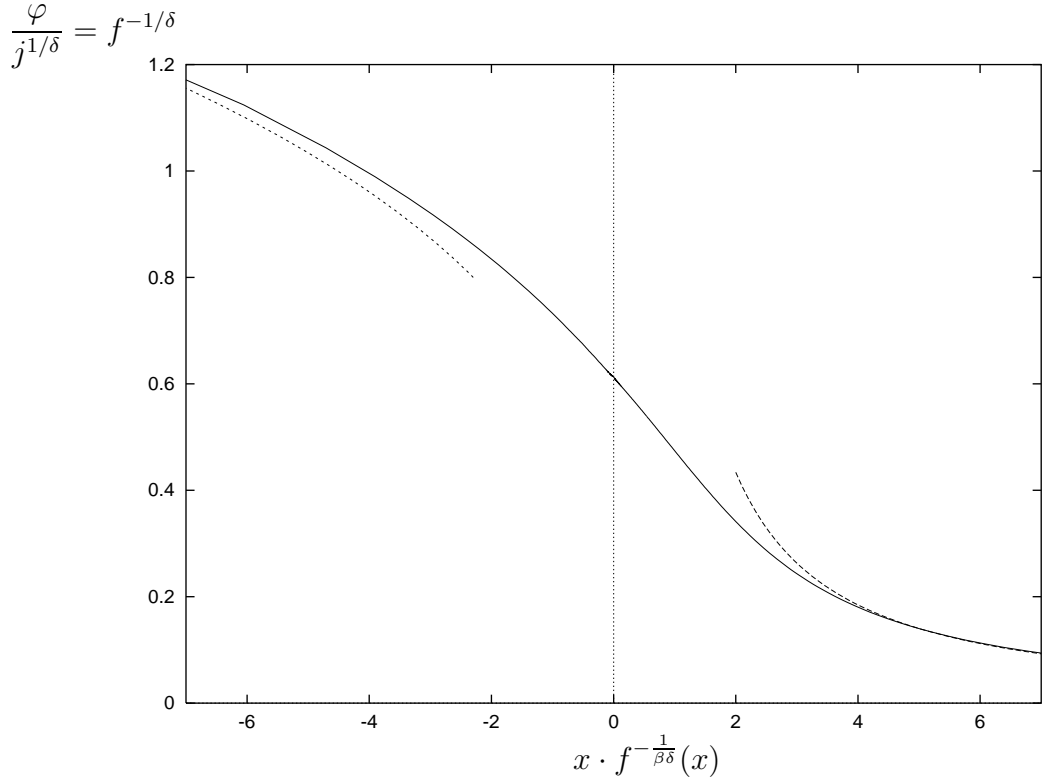
}}
}
%\put(0.,0.){\framebox(10.,10.)}
\end{picture}
\end{center}
\caption[]{\label{Z2Widom}
\footnotesize
Widom scaling function $f(x)$ of the $(d=3)$-Ising model. (
The present curve is generated for a quartic
short distance potential $U_{\Lambda}$ with $\lambda_{\Lambda}/\Lambda=5$).
The dashed lines indicate extrapolations of the limiting behaviour as given
by the critical exponents.}
\end{figure}

In the universal critical region the expectation value of $\vp$
for very small $|\delta m_{\Lambda}^2|$ and sources $j$
(with \mbox{$\partial U/\partial\vp(\vp_j)=j$}) can be found
from the Widom scaling form of the equation of state
\bea
   \ds \frac{\partial}{\partial\vp}U(\vp) & = & \ds |\vp|^{\delta} f(x)
       \cdot\Lambda^{\frac{5-\delta}{2}}
   \label{IsingWidomSklform}
\eea
\bea
   \ds x & = & \ds \frac{\delta m_{\Lambda}^2}{|\vp|^{1/\beta}}
                   \cdot \Lambda^{\frac{1}{2\beta}-2}.
\eea
Our results for the scaling function $f(x)$ are shown in fig. \ref{Z2Widom},
 together with the asymptotic behaviour (dashed lines) as dictated by the
critical exponents and amplitudes:
\bea
   \ds \lim_{x\to 0} f(x) = D & , & \ds f(x\!=\!-B^{-1/\beta})=0,   \nnn
   \ds \lim_{x \to \infty} f(x) = (C^{+})^{-1} x^{\gamma} & , &
   \ds \lim_{x \to -\infty} f(x)=(C^{-})^{-1}(-x)^{\gamma}  \nnn
   \label{Grzfll2}
\eea
where
\bea
   \ds (C^{-})^{-1} = B^{\delta-1-1/\beta}\frac{1}{\beta}
       f^{\prime}(-B^{-1/\beta}) & , &
   \ds f^{\prime}(x)=\frac{\beta}{x}\cdot\left(\delta f(x)
     -  \frac{\partial^2 U}{\partial\vp^2}\vp^{1-\delta}\right).
\eea
We have verified the scaling form of the equation of state explicitly by
starting with a large variety of initial conditions (e.g. different
$\lambda_{\Lambda}$).

% Bild: Widom-Skalenfunktion f(x) des Ising-Modells

A different useful parameterization of the critical equation of state can be
given in terms of nonlinearly rescaled fields $\hat{\vp}_R$, using a
$\vp$-dependent wave function renormalization $Z(\vp)=Z_{k\to 0}(\vp)$,
\bea
   \ds \hat{\vp}_R & = & \ds Z(\vp)^{1/2}\vp.
\label{rescaled_field}
\eea
Our numerical results can be presented in terms of a fit to the universal
function
\bea
   \ds \hat{F}(\hat{s})=m_R^{-5/2}\frac{\partial U}{\partial\hat{\vp}_R} & , &
   \ds \hat{s}=\frac{\hat{\vp}_R}{m_R^{1/2}}
              =\left(\frac{Z(\vp)\vp^2}{m_R}\right)^{1/2},  \nnn
%   \ds \hat{G}(\hat{s}):=\frac{\hat{U}_0^{\prime}(\hat{\vp}_R)}{m_R^{5/2}} & &
%   \ds \hat{s}=\frac{\hat{\vp}_R}{m_R^{1/2}}.
\eea
\bea
  \ds  \hat{F}_{Fit}(\hat{s}) & = & \left(a_0\hat{s}+a_1\hat{s}^3+a_2\hat{s}^5
        +a_3\hat{s}^7\right)\cdot f_{\alpha}(\hat{s})
        +(1-f_{\alpha}(\hat{s}))\cdot a_4\hat{s}^5.
\eea
The factors $f_{\alpha}$ and $(1-f_{\alpha})$ interpolate between a polynomial
expansion and the asymptotic behaviour for large arguments. We use
\bea
  \ds f_{\alpha}(x) & = & \ds \alpha^{-2} x^2\cdot
           \frac{\exp(-\frac{x^2}{\alpha^2})}{1-\exp(-\frac{x^2}{\alpha^2})}.
\eea
A similar fit can be given for
\bea
  \ds \tilde{z}(s) = \frac{\lim_{k\to 0} Z_k(\vp)}{Z_0} & , &
   \ds s=\frac{\vp_R}{m_R^{1/2}}
        =\left(\frac{Z_0\vp^2}{m_R}\right)^{1/2}=\tilde{z}^{-1/2}\hat{s},
\eea
\bea
  \ds  \tilde{z}_{Fit}(s) & = & \left(b_0+b_1 s^2+b_2 s^4+b_3s^6
        +b_4 s^8\right)\cdot f_{\beta}(s)
        +(1-f_{\beta}(s))\cdot b_5 |s|^{-\frac{2\eta}{1+\eta}}.
\eea
In the symmetric phase one finds (with $\eta=0.0467$) $\alpha=1.012$,  
$a_0=1.0084$, $a_1=3.1927$, $a_2=9.7076$, $a_3=0.5196$, $a_4=10.3962$
and $\beta=0.5103$, $b_0=1$, $b_1=0.3397$, $b_2=-0.8851$, \mbox{$b_3=0.8097$},
 $b_4=-0.2728$, $b_5=1.0717$, whereas the fit parameters for
the phase with spontaneous symmetry breaking are
$\alpha=0.709$, $a_0=-0.0707$, $a_1=-2.4603$, \mbox{$a_2=11.8447$},
$a_3=-1.3757$, $a_4=10.2115$ and
$\beta=0.486$, $b_0=1.2480$, $b_1=-1.4303$, $b_2=2.3865$, $b_3=-1.7726$,
$b_4=0.4904$, $b_5=0.8676$
(our fit parameters are evaluated for this phase for
$(\partial^2 U/\partial\vp_R^2)(\vp_{R,max})/k^2=-0.99$).
One observes that the coefficients $a_2$ and $a_4$ are large and of comparable
size. A simple polynomial form
$\hat{F}=\tilde{a}_0 \hat{s} + \tilde{a}_1\hat{s}^3+ \tilde{a}_2\hat{s}^5$
is not too far from the more precise result. We conclude that in terms of
the rescaled field $\hat{\vp}_R$ (\ref{rescaled_field}) the potential is
almost a polynomial $\vp^6$-potential.

% Bild: Universelle reskalierte Wellenfunktionsrenormierung Z/Z0(s) in
%       symmetr. Phase
\begin{figure}[h]
\unitlength1.0cm
%\begin{center}
%\begin{picture}(6.5,4.5)
\begin{picture}(6.5,6.5)
\put(3.5,-0.5){$s=\frac{\vp_R}{m_R^{1/2}}$ }
%\put(0.,3.){$\tilde{z}_{k\to 0}(s)$}
\put(1.2,4.5){$\tilde{z}$}
\put(0.,0.){
\epsfysize=7.5cm
\epsfxsize=5.5cm
\rotate[r]{\epsffile{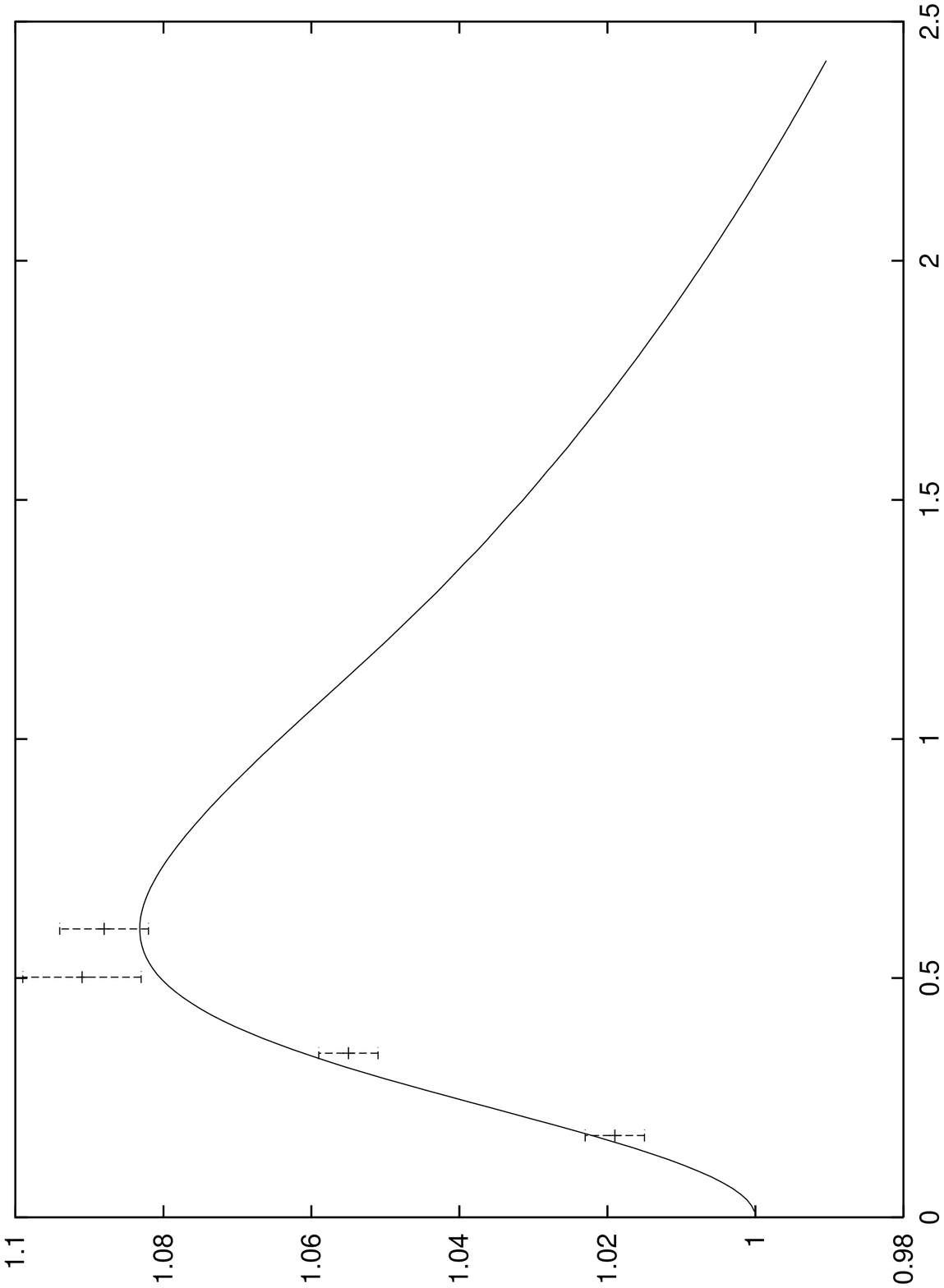   % univZsym.ps
}}
}
%\put(0.,0.){\framebox(10.,10.)}
\end{picture}
%\end{center}
%\caption[]{\label{univZsym}
%\footnotesize
%Universelle reskalierte Wellenfunktionsrenormierung $\tilde{z}_{k\to 0}$
%in der symmetrischen Phase des
%$(d=3)$-Ising-Modells.}
%\end{figure}

% Bild: Universelle reskalierte Wellenfunktionsrenormierung Z/Z0(s) in
%       gebr. Phase
%\begin{figure}[h]
%\unitlength1.0cm
%\begin{center}
\begin{picture}(6.,0.)
\put(11.5,0.){$s=\frac{\vp_R}{m_R^{1/2}}$ }
%\put(9.,3.5){$\tilde{z}_{k\to 0}(s)$}
\put(11.5,3.5){$\tilde{z}_k$}
\put(9.5,2.8){\footnotesize $c=-0.9$}
\put(10.6,5.){\footnotesize $c=-0.99$}
%\put(-0.5,0.){
\put(8.5,0.5){
\epsfysize=7.5cm
\epsfxsize=5.5cm
\rotate[r]{\epsffile{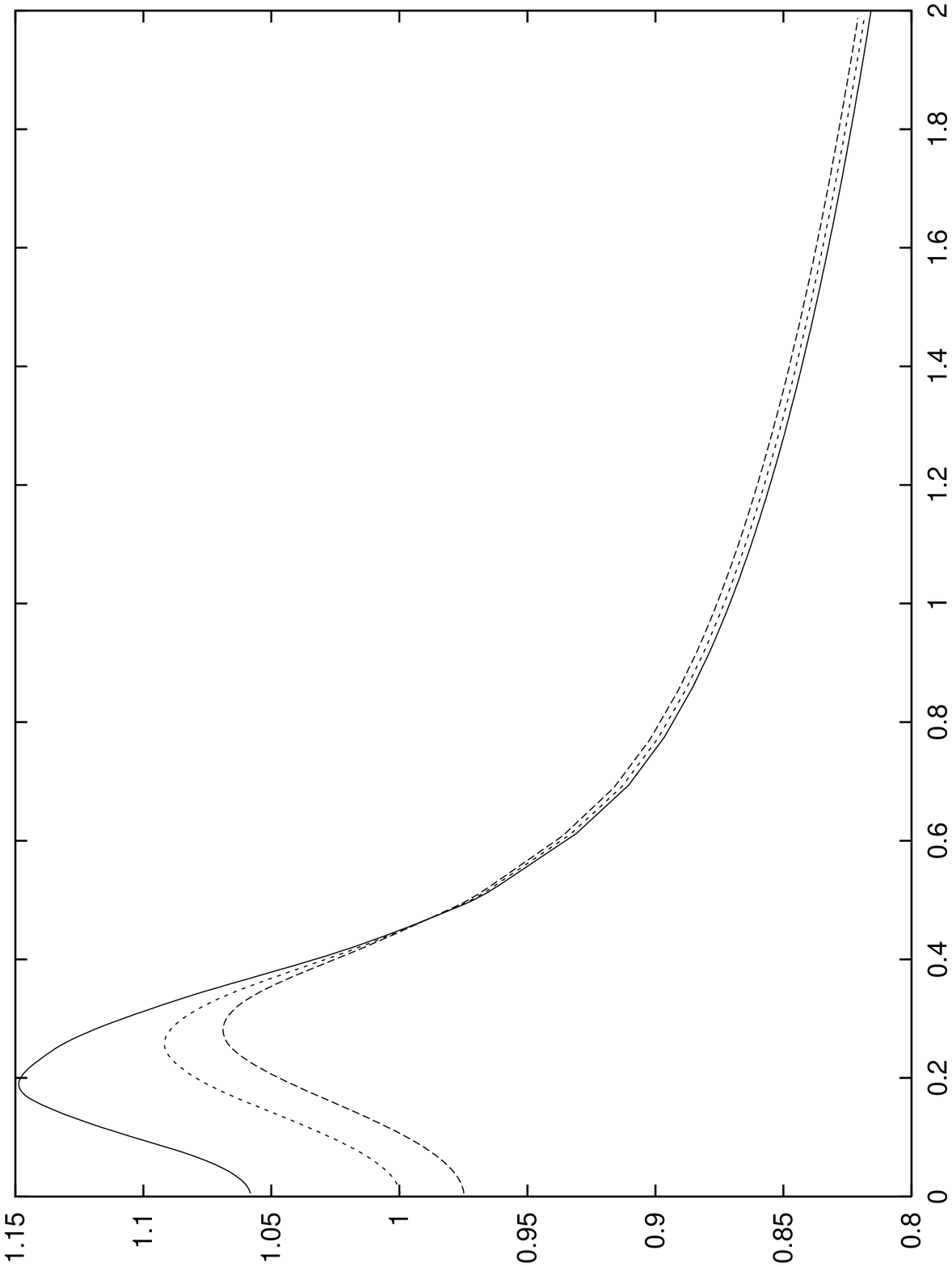
}}
}
%\put(0.,0.){\framebox(10.,10.)}
\end{picture}
%\end{center}
\caption[]{\label{univz}
\footnotesize
Universal rescaled wave function renormalization $\tilde{z}$
in the symmetric and ordered phase of the $(d\!=\!3)$-Ising model. In the
low temperature phase the plots are for different $k$ with
$\frac{1}{k^2}\frac{\partial^2 U}{\partial \vp_R^2}(\vp_{R,max})\!=\!c$
($c=-0.9,-0.95,-0.99$). Here $\vp_{R,max}$ is the location of the
local maximum of the potential in the inner (non-convex) region.
In the graph for the high temperature phase we have inserted Monte
Carlo results by M. Tsypin (private communication).
}
\end{figure}

In fig. \ref{univz} we show $\tilde{z}$ as a function of $s$ both for
the symmetric and ordered phase. Their shape is similar to the scaling
solution found in \cite{Morris94}.
Nevertheless, the form of $\tilde{z}$ for $k=0$ which expresses directly
information about the physical system should not be confounded with the
scaling solution which  depends on the particular infrared cutoff.
For the low temperature phase one sees
the substantial dependence of $\tilde{z}_k$ on the infrared cutoff $k$ for
small values $s<s_0$. Again this corresponds to the ``inner region'' between
the origin ($s\!=\!0$) and the minimum of the potential ($s_0\!=\!0.449$) where
the potential finally becomes convex for $k\to 0$.

Knowledge of $U$ and $\tilde{z}$ permits the computation of the (renormalized)
propagator for low momenta with arbitrary sources $j$. It is given by
\bea
  \ds G(q^2)=\left(\frac{\partial^2 U(\vp_R)}{\partial \vp_R^2}
             + \tilde{z}(\vp_R)q^2\right)^{-1}
\eea
for $\tilde{z}q^2 \lta \partial^2 U/\partial \vp_R^2$. Here $\vp_R$
obeys $\partial U/\partial\vp_R = Z_0^{-1/2} j$. We emphasize that the
correlation length
$\xi(\vp_R)=\tilde{z}^{1/2}(\vp_R)(\partial^2 U/\partial\vp_R^2)^{-1/2}$
at given source $j$ requires information about $\tilde{z}$.
For the gas-liquid transition $\xi(\vp_R)$ is directly connected to the
density dependence of the correlation length. For magnets, it expresses the
correlation length as a function of magnetization. The factor
$\tilde{z}^{1/2}$ is often omitted in other approaches.

\begin{figure}[h]
\unitlength1.0cm
\begin{center}
\begin{picture}(13.,9.)
\put(10.5,2.8){\footnotesize (1)}
\put(11.,5.5){\footnotesize(4)}
\put(9.9,7.){\footnotesize(3)}
\put(11.1,7.4){\footnotesize(2)}
\put(11.1,6.8){\footnotesize(5)}
\put(6.5,-0.5){$\tilde{s}$}
\put(0.7,8.){$\ds\frac{F(\tilde{s})}{c_F}$}
%\put(7.2,1.){\vector(-1,1){0.4}}

\put(-0.5,0.){
\epsfysize=13.cm
\epsfxsize=9.cm
\rotate[r]{\epsffile{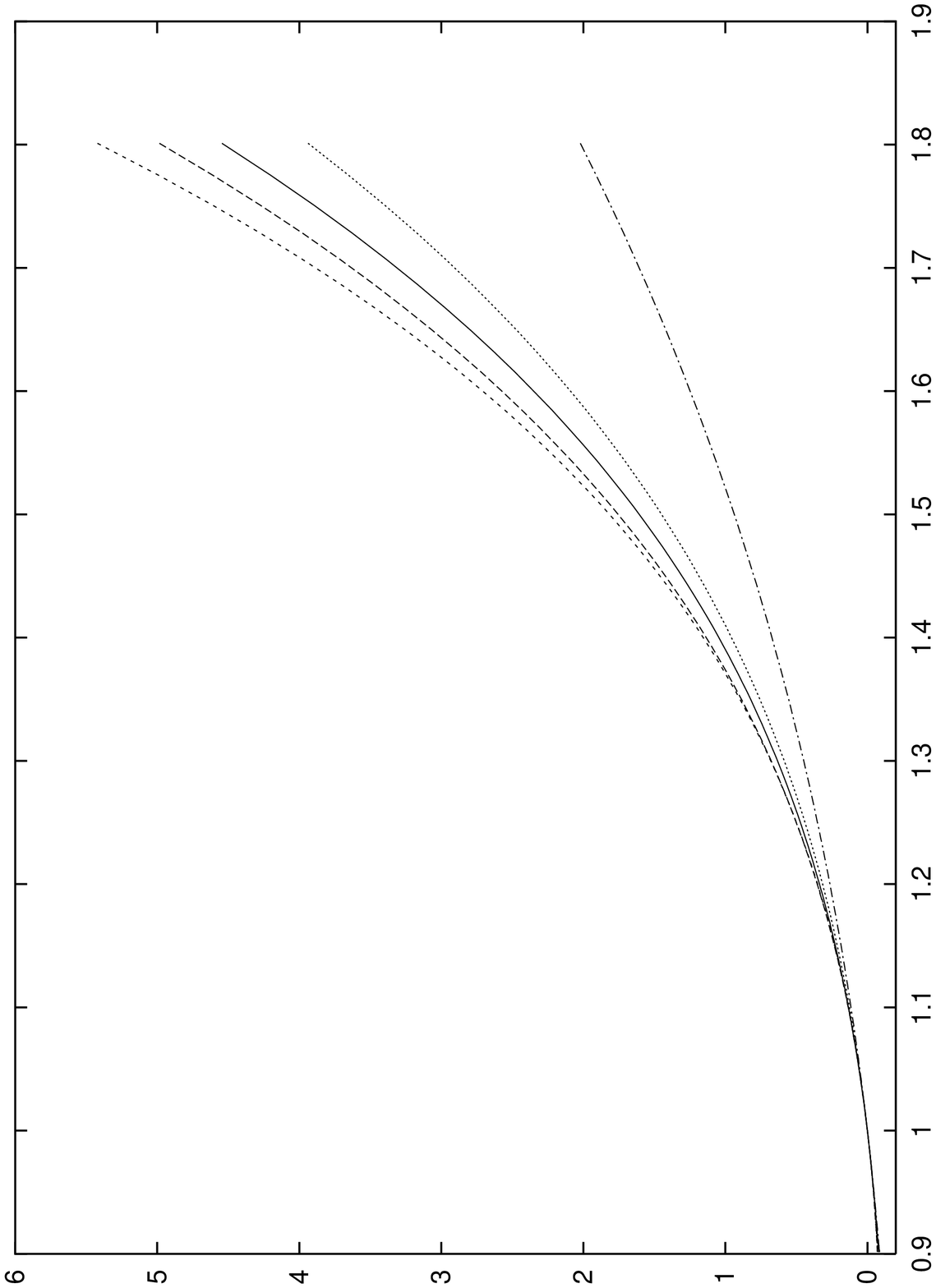
}}
}
%\put(0.,0.){\framebox(10.,10.)}
\end{picture}
\end{center}
\caption[]{\label{compEquofStat}
\footnotesize
The critical equation of state in the ordered phase. \protect \\
(1) mean field approximation. \protect \\
(2) effective average action with uniform wave function renormalization
    \protect\cite{BergTetWet97}. \protect \\
(3) Monte Carlo simulation \cite{Tsypin96}. \protect \\
(4) resummed $\epsilon$-expansion in $O(\epsilon^3)$, five loop perturbative
    expansion and high temperature series
    \cite{GuidaZinn96}. \protect \\
(5) present work.}
\end{figure}

\begin{figure}[h]
\unitlength1.0cm
\begin{center}
\begin{picture}(13.,9.)
\put(11.4,8.2){\footnotesize (1)}
\put(10.8,7.5){\footnotesize (2)}
\put(11.3,7.4){\vector(1,-1){0.45}}
\put(10.5,7.1){\footnotesize (3)}
\put(10.9,7.0){\vector(1,-1){0.59}}
\put(10.,6.){\footnotesize(4)}
\put(10.5,6.0){\vector(1,-1){0.65}}
\put(11.15,4.5){\footnotesize (5),(6)}
\put(11.7,4.8){\vector(-1,2){0.33}}
\put(10.6,3.4){\footnotesize (7)}
\put(10.8,3.7){\vector(-1,1){0.4}}
\put(6.5,-0.5){$\tilde{s}$}
\put(1.,8.){$\ds F(\tilde{s})$}

\put(-0.5,0.){
\epsfysize=13.cm
\epsfxsize=9.cm
\rotate[r]{\epsffile{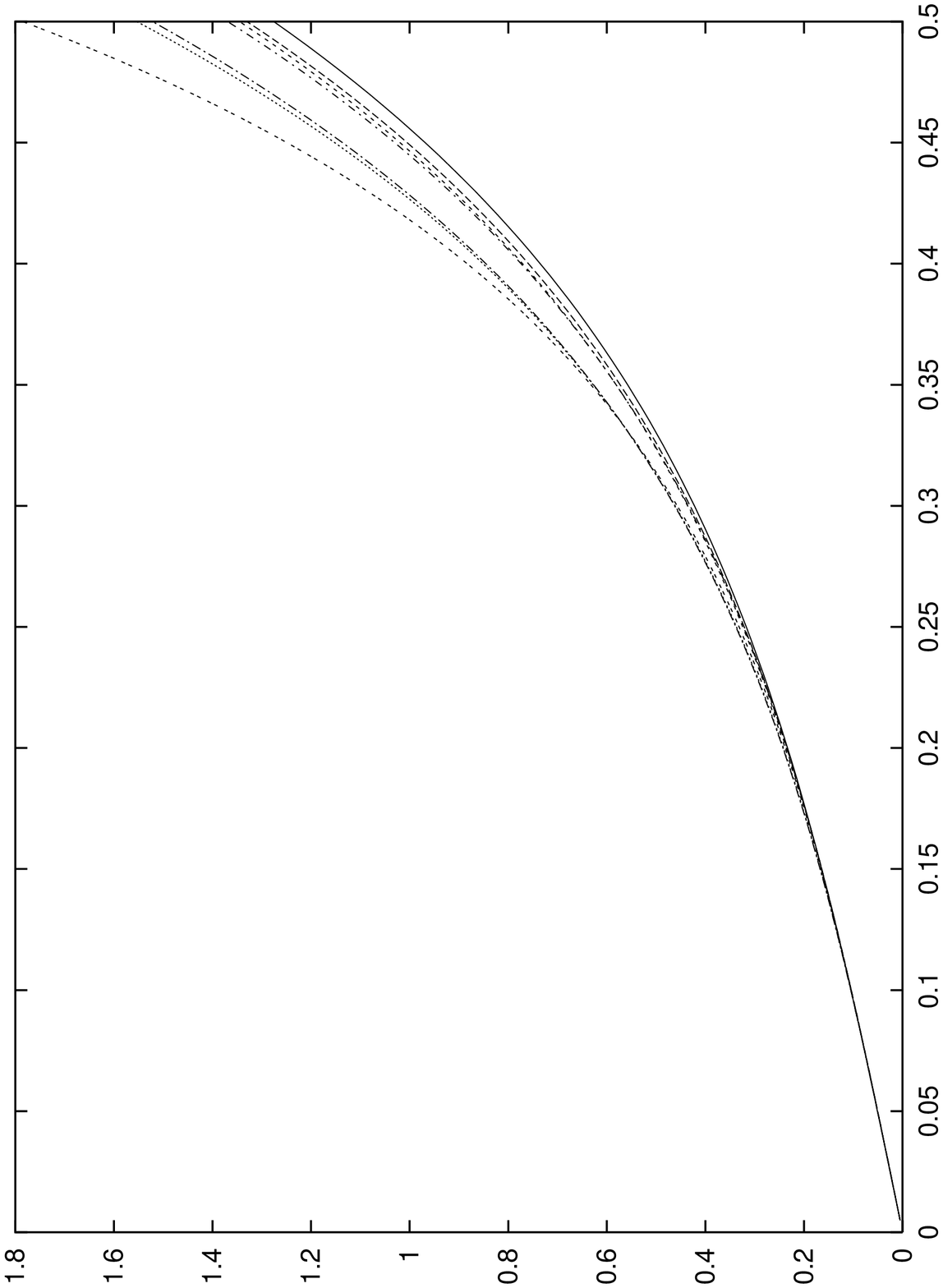
}}
}
%\put(0.,0.){\framebox(10.,10.)}
\end{picture}
\end{center}
\caption[]{
\footnotesize
The critical equation of state in the symmetric phase. \protect \\
(1) Monte Carlo simulation \protect\cite{KimLandau96},
(2) $\epsilon$-expansion \protect\cite{GuidaZinn96}. \protect \\
(3)  effective average action with uniform wave function renormalization
    \protect\cite{BergTetWet97}. \protect \\
(4) Monte Carlo simulation \protect\cite{Tsypin94},
(5),(6) high temperature series
    \protect\cite{ButeraComi96},\protect\cite{ZinnLaiFisher96}.
    \protect \\
(7) present work.
\label{compEquofStatSym}
}
\end{figure}

Critical equations of state for the Ising model have been computed earlier
with several methods.
They are compared
with our result for the phase with spontaneous symmetry breaking in
\mbox{fig. \ref{compEquofStat}} and for the symmetric phase in
\mbox{fig. \ref{compEquofStatSym}}.
For this purpose we use $F(\tilde{s})=m_R^{-5/2}\partial U/\partial\vp_R$
with $\tilde{s}=\frac{\vp_R}{\vp_{0R}}$ in the phase with spontaneous symmetry
breaking (note $\tilde{s}\sim s$). The constant $c_F$ is adapted such that
$\frac{1}{c_F}\frac{\partial F}{\partial \tilde{s}}(\tilde{s}\!=\!1)\!=\!1$.
In the symmetric phase we take instead
$\tilde{s}\!=\!\frac{\vp_R}{m_R^{5/2}}$ such that
$\frac{\partial F}{\partial \tilde{s}}(\tilde{s}\!=\!0)\!=\!1$.
One expects for large $\tilde{s}$ an inaccuracy of our results due to the
error in $\eta$.

% Bild: Vergleich der universellen Zustandsgleichungen in der gebrochenen
% Phase

% Bild: Vergleich der universellen Zustandsgleichungen in der symmetrischen
% Phase

\begin{table} [h] \centering
{ \begin{tabular}
{|l|c|c|c|c|c|c|c|c|}
\hline
   & $m_{\Lambda,crit}^2$ & $C^{+}$ & D \\
\hline
\hline
 $\lambda_{\Lambda}=0.1$ & $-6.4584\cdot 10^{-3}$ & $0.1655$ & $5.3317$ \\
\hline
 $\lambda_{\Lambda}=1$   & $-5.5945\cdot 10^{-2}$ & $0.485$ & 7.506 \\
\hline
 $\lambda_{\Lambda}=5$ & $-0.22134$ & $1.033$ & $10.213$ \\
\hline
 $\lambda_{\Lambda}=20$ & $-0.63875$ & $1.848$ & $16.327$ \\
\hline

\end{tabular}}

%\hspace*{\fill}
\caption{\footnotesize  The critical values
$m_{\Lambda,crit}^2$ and the non-universal amplitudes $C^{+}$, $D$ as a
function of the quartic short distance coupling $\lambda_{\Lambda}$
(all values expressed in units of $\Lambda$).
Other non-universal amplitudes can be calculated from the universal quantities
of table \ref{IsingUnivAmpTab}.
\label{nonunivtable} }
\end{table}

In summary of this section we may state that the non-perturbative flow
equations in second order in a derivative expansion lead to a critical equation
of state which is well compatible with high order expansions within other
methods. In addition, it allows to establish an explicit connection between
the parameters appearing in the microscopic free energy $\Gamma_{\Lambda}$ and
the universal long distance behaviour. For a quartic polynomial potential
this involves in addition to the non-universal amplitudes the
value of $m_{\Lambda,crit}^2$. We have listed these quantities for different
values
of $\lambda_{\Lambda}$ in table \ref{nonunivtable}. Finally, the temperature
scale is
established by $S=\partial m_{\Lambda}^2/\partial T\left.\right|_{T_c}$.

% ****************************************************************************
\vspace{2cm}
\sect{Equation of state for first order transitions \label{cubicmodel}}

Our method is not restricted to a microscopic potential with discrete
$Z_2$-symmetry. The numerical code works for arbitrary initial potentials.
We have investigated the polynomial potential (\ref{microsU}) with
$\gamma_{\Lambda}\neq0$. The numerical solution of the flow equations
(\ref{dtu}),(\ref{dtz}) shows the expected first order
transition (in case of vanishing linear term $j$).  Quite
generally, the universal critical equation of state for first order transitions
will depend on two scaling parameters (instead of one for second order
transitions) since the jump in the order parameter or the mass introduces
a new scale. The degree to which universality applies depends on the
properties of a given model and its parameters (see
\cite{numsolreference,BergTetWet97} for a detailed discussion).
For a $\vp^4$-model with cubic term
(\ref{microsU}) one can relate the equation of state to
the Ising model by an appropriate mapping. This allows us to
compute the universal critical equation of state for arbitrary first order
phase transitions in the Ising universality class from the critical equation
of state for the second order phase transition in the Ising model.
For other universality classes
a simple mapping to a second order equation of state
is not always possible - its existence is particular to the present model.

By a variable shift
\bea
 \ds  \sigma & = & \ds \vp+\frac{\gamma_{\Lambda}}{3\lambda_{\Lambda}}
 \label{varshift}
\eea
we can bring the short distance potential (\ref{microsU}) into the form
\bea
   \ds U_{\Lambda}(\sigma) & = &
   \ds -J_{\gamma}\sigma + \frac{\mu_{\Lambda}^2}{2}\sigma^2
   + \frac{\lambda_{\Lambda}}{8} \sigma^4 + c_{\Lambda}
\eea
with
\bea
  \ds J_{\gamma} & = &
  \ds \frac{\gamma_{\Lambda}}{3\lambda_{\Lambda}}m_{\Lambda}^2
      -\frac{\gamma_{\Lambda}^3}{27\lambda_{\Lambda}^2} \nnn
  \ds \mu_{\Lambda}^2 & = &
  \ds  m_{\Lambda}^2 -\frac{\gamma_{\Lambda}^2}{6\lambda_{\Lambda}}.
\eea
We can now solve the flow equations in terms of $\sigma$ and reexpress the
result in terms of $\vp$ by eq. (\ref{varshift}) at the end. The exact flow
equation (\ref{ERGE}) does not involve the linear term $\sim J_{\gamma}\sigma$
on the right hand side (also the constant $c_{\Lambda}$ is irrelevant).
Therefore the effective potential ($k=0$) is given by
\bea
  \ds U=U^{Z_2}(\sigma)-J_{\gamma}\sigma + c_{\Lambda}& = &
  \ds U^{Z_2}(\vp+\frac{\gamma_{\Lambda}}{3\lambda_{\Lambda}})
      -(\vp+\frac{\gamma_{\Lambda}}{3\lambda_{\Lambda}}) J_{\gamma}
      + c_{\Lambda},
  \label{shiftU}
\eea
where $U^{Z_2}$ is the effective potential of the Ising type model with
quartic coupling $\lambda_{\Lambda}$ and mass term $\mu_{\Lambda}^2$. The
equation of state $\partial U/\partial\vp=j$ or, equivalently
\bea
  \ds \frac{\partial U^{Z_2}}{\partial\vp}\left.
           \right|_{\vp+\frac{\gamma_{\Lambda}}{3\lambda_{\Lambda}}} & = &
  \ds j + J_{\gamma},
  \label{shifteqofstat}
\eea
is therefore known explicitly for arbitrary $m_{\Lambda}^2$, $\gamma_{\Lambda}$
and $\lambda_{\Lambda}$ (cf. eq. (\ref{IsingWidomSklform}) for the
universal part). This leads immediately to the following conclusions:
\begin{itemize}
 \item[i)] First order transitions require that the combination $U(\vp)-j\vp$
           has two degenerate minima. This happens for $J_{\gamma}+j=0$ and
           $\mu_{\Lambda}^2<\mu_{\Lambda,crit}^2$ or
           \bea
           \ds j & = & \ds -\frac{\gamma_{\Lambda}}{3\lambda_{\Lambda}}
                \left(m_{\Lambda}^2-\frac{\gamma_{\Lambda}^2}
                {9\lambda_{\Lambda}}\right)
           \label{jcrit}
           \eea
           \bea
           \ds m_{\Lambda}^2 & < & \ds \mu_{\Lambda,crit}^2
                + \frac{\gamma_{\Lambda}^2}{6\lambda_{\Lambda}}.
           \eea
           Here $\mu_{\Lambda,crit}^2$ is the critical mass term of the
           Ising model.
 \item[ii)] The boundary of this region for
            \bea
            \ds m_{\Lambda}^2 & = &
            \ds \mu_{\Lambda,crit}^2+\frac{\gamma_{\Lambda}^2}
                {6\lambda_{\Lambda}}
            \label{2ndorderline}
            \eea
            is a line of second order phase transitions with vanishing
            renormalized mass or infinite correlation length.
\end{itemize}
For $j=0$ (e.g. magnets with polynomial potential in absence of external
fields) the equations (\ref{jcrit}), (\ref{2ndorderline}) have the solutions
\bea
  \ds \gamma_{\Lambda;1}=0 & , &
  \ds \gamma_{\Lambda;2,3}=\pm
       (-18\lambda_{\Lambda}\mu_{\Lambda,crit}^2)^{1/2}.
\eea
The second order phase transition for $\gamma_{\Lambda}\neq 0$ can be described
by Ising models for shifted fields $\sigma$.
For a given model, the way how a phase transition line is crossed as the
temperature is varied follows from the temperature dependence  of $j$,
 $m_{\Lambda}^2$, $\gamma_{\Lambda}$ and $\lambda_{\Lambda}$. For the
gas-liquid transition both $j$ and $m_{\Lambda}^2$ depend on $T$.

In the vicinity of the boundary of the region of first order transitions the
long range fluctuations play a dominant role and one expects universal
critical behaviour. The detailed microscopic physics is only reflected in two
non-universal amplitudes. One reflects the relation between the renormalized
and unrenormalized fields as given by $Z_0$. The other is connected to the
renormalization factor for the mass term. Expressed in terms of renormalized
fields and mass the potential $U$ looses all memory about the microphysics.

The critical equation of
state of the non-symmetric model ($\gamma_{\Lambda}\neq 0$) follows from
the Ising model (4.5). With
$\frac{\partial U^{Z_2}}{\partial\vp}\left.\right|_{\vp}=|\vp|^{\delta}f(x)$,
the scaling form of the equation of state
$j=\frac{\partial U}{\partial \vp}$ for the model with cubic coupling can be
written as
\bea
  \ds j  & = & \ds |\vp+\frac{\gamma_{\Lambda}}{3\lambda_{\Lambda}}|^{\delta}
       f(x)
      - \left(\frac{\gamma_{\Lambda}}{3\lambda_{\Lambda}}\mu_{\Lambda,crit}^2
          + \frac{\gamma_{\Lambda}^3}{54\lambda_{\Lambda}^2}\right)
      - \frac{\gamma_{\Lambda}}{3\lambda_{\Lambda}}\delta\mu_{\Lambda}^2,
\eea
where $\ds x=\frac{\delta\mu_{\Lambda}^2}
{|\vp+\frac{\gamma_{\Lambda}}{3\lambda_{\Lambda}}|^{1/\beta}}$ and
$\delta\mu_{\Lambda}^2=m_{\Lambda}^2-\frac{\gamma_{\Lambda}^2}
{6\lambda_{\Lambda}}-\mu_{\Lambda,crit}^2$.
One may choose
\bea
  \ds y & = & \ds
      \frac{\gamma_{\Lambda}}{3\lambda_{\Lambda}}
      \left(\mu_{\Lambda,crit}^2+\frac{\gamma_{\Lambda}^2}{18\lambda_{\Lambda}}
        + \delta\mu_{\Lambda}^2\right)
      |\varphi+\frac{\gamma_\Lambda}{3\lambda_\Lambda}|^{-\delta}
\eea
as the second scaling variable. For small symmetry breaking cubic coupling
$\gamma_{\Lambda}$ one notes $y\sim\gamma_{\Lambda}$. The scaling form of the
equation of state for the non-symmetric model reads
\bea
  \ds  j & = & \ds
       |\varphi+\frac{\gamma_\Lambda}{3\lambda_\Lambda}
       |^{\delta}\left\{ f(x) - y \right\}.
\eea
This universal form of the equation of state is relevant for a large class of
microscopic free energies, far beyond the special polynomial form used for
its derivation.

It is often useful to express the universal equation
of state in terms of renormalized fields and masses. We use the variables
\bea
  \ds \tilde{s}=\frac{\vp_R}{\vp_{0R}} & ; &
  \ds v=\frac{m_R}{m_R^{Z_2}},
\eea
where $m_R=\left(\frac{\partial^2 U}{\partial\vp_R^2}\left.\right|_{\vp_{0R}}
\right)^{1/2}$ is the renormalized mass at the minimum
$\vp_{0R}$ of $U(\vp_R)$ whereas
$m_R^{Z_2}$ is the renormalized mass at the minimum of the corresponding
$Z_2$-symmetric effective potential obtained for vanishing cubic coupling
$\gamma_{\Lambda}=0$. Then the critical temperature corresponds to $v=1$.
In this parameterization the universal properties of the equation of state
for the Ising type first order transition can be compared with transitions
in other models - e.g. matrix models \cite{numsolreference} - where no simple
mapping to a second order phase transition exists.

\begin{figure}[h]
\unitlength1.0cm
\begin{center}
\begin{picture}(13.,9.)
\put(6.5,-0.5){$\tilde{s}=\frac{\vp_R}{\vp_{0R}}$ }
\put(1.2,7.){$G(\tilde{s},v)$}
\put(7.7,3.8){\footnotesize $v=1$}
\put(9.7,0.6){\footnotesize $v=1.195$}
\put(-0.5,0.){
\epsfysize=13.cm
\epsfxsize=9.cm
\rotate[r]{\epsffile{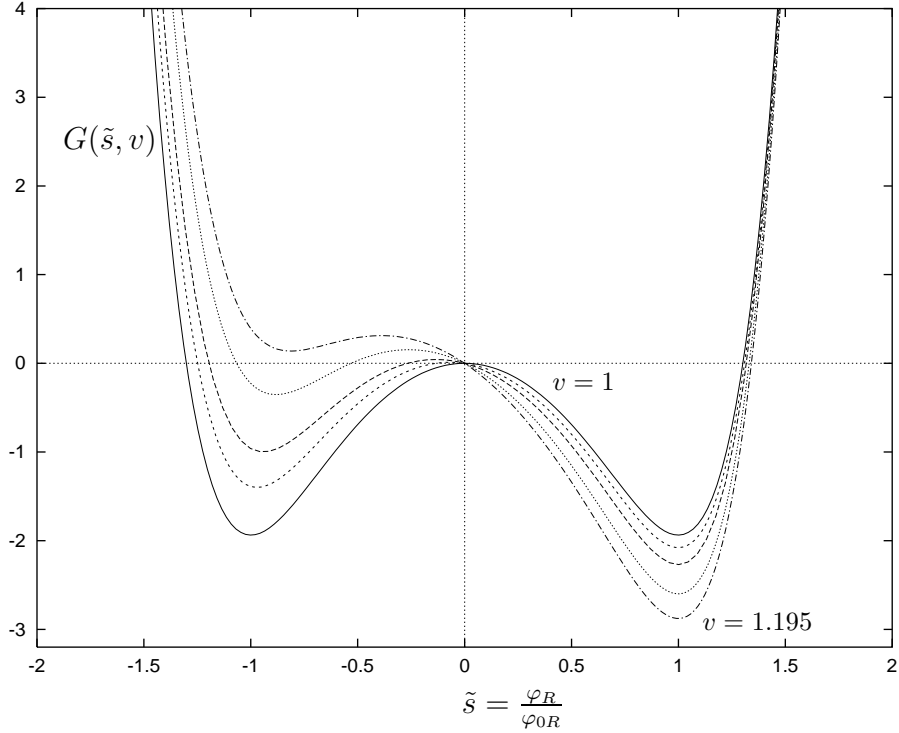           %  GUkub.ps
}}
}
%\put(0.,0.){\framebox(10.,10.)}
\end{picture}
\end{center}
\caption[]{
\footnotesize $G(\tilde{s},v)$ for $v=1$, $v=1.037$, $v=1.072$, $v=1.137$ and 
$v=1.195$.}
\label{univGkub}
\end{figure}

A convenient universal function $G(\tilde{s},v)$ for weak first order
transitions can be defined as
\bea
  \ds G(\tilde{s},v):=\frac{U(\vp_R)}{\vp_{0R}^6}.
\eea
We plot $G(\tilde{s},v)$ in fig. \ref{univGkub} as a function of $\tilde{s}$
for different values of $v$. For the present model all information necessary
for a universal description of first order phase transitions is already
contained in eqs. (\ref{shiftU}) or (\ref{shifteqofstat}).
The function $G(\tilde{s},v)$ can serve, however, for a comparison with other
models. At this place we mention that we
have actually computed the potential $U$ both by solving the flow equations
with initial values where $\gamma_{\Lambda}\neq 0$ and by a shift from the
Ising model results. We found good agreement between the two approaches.

% Bild:  Universelle Funktionen des allgemeinen Modells

% ***********************************************************************
\sect{Discussion}

In conclusion, we have employed non-perturbative flow equations in order
to compute explicitly the equation of state. We have first studied models
where the microscopic free energy can be approximated by a polynomial
approximation with terms up to quartic order. This covers second order
as well as first order transitions, both for the universal and non-universal
features. The same method can be used away from the critical hypersurface,
allowing therefore for an explicit connection between critical and
non-critical observations.

The ability of the method to deal also with a microscopic free energy which
is not of a polynomial form is demonstrated by a particular example,
namely the equation of state for carbon dioxide.
In the vicinity of the endpoint of the critical line we can
give an explicit formula for the free energy density $U(n,T)T$.
Using the fits (3.15), ((3.18), for $\hat F(\hat s)$ and
$\tilde z(s)$ one finds\footnote{Note that
$n_*$ is somewhat different form $\hat n$ and therefore
$\varphi_R$ is defined slightly different from eq. (1.7).
This variable shift (similar to (4.1) reflects the fact
that eq. (1.6) contains higher than quartic interactions
and cannot be reduced to a $\varphi^4$-potential even for
$\gamma_\Lambda=0$.}
\be\label{5.1}
U(n,T)=U^{Z_2}(\hat\varphi_R(n,T),\ m_R(T))
+J(T)(n-n_*)-K(T)\ee
\be\label{5.2}
U^{Z_2}=\frac{1}{2}\tilde a_0m^2_R\hat\varphi_R^2+\frac{1}{4}
\tilde a_1m_R\hat\varphi^4_R+\frac{1}{6}\tilde a_2\hat\varphi^6_R\ee
with $\tilde a_i\approx a_i$ and
\ben\label{5.3}
\hat\varphi_R(n,T)&=&\tilde z\left(\frac{\varphi_R(n,T)}
{m^{1/2}_R(T)}\right)\varphi_R(n,T)\nonumber\\
\varphi_R(n,T)&=&H_{\pm}\left|\frac{T-T_*}{T_*}\right|
^{-\eta\nu}(n-n_*)\nonumber\\
m_R(T)&=&\xi_{T\pm}\left|\frac{T-T_*}{T_*}\right|^\nu\een
The two non-universal functions $J(T)$ and $K(T)$ enter in
the determination of the chemical potential and the critical line.

In particular, the nonuniversal amplitudes governing the behaviour near the
endpoint of the critical line can be extracted from the equation of state:
In the vicinity of the endpoint we find for $T=T_*$
\bea
 \ds \rho_{>}-\rho_* = \rho_*-\rho_{<} & = &
 \ds D_p^{-1}\left(\frac{| p-p_*|}{p_*} \right) ^{1/\delta}
\eea
with $D_p=2.8\: g^{-1} cm^3$,
where $\rho_{>}>\rho_{*}$ and $\rho_{<}<\rho_*$
refer to the density in the high and low
density region respectively. At the critical temperature $T_c < T_{*}$ and
pressure $p_c<p_*$ for a first order transition one finds for the
discontinuity in the density between the liquid $(\rho_l)$ and gas
$(\rho_g)$ phase
\bea
 \ds \Delta \rho = \rho_l - \rho_g & = &
 \ds B_p\left(\frac{p_*-p_c}{p_*}\right)^{\beta} =
     B_T\left(\frac{T_*-T_c}{T_*}\right)^{\beta}
\eea
with $B_p=0.85 \: g cm^{-3}$, $B_T=1.5 \: g cm^{-3}$.
This relation also defines the slope of the critical line near the endpoint.

There is no apparent limitation for the use of the flow equation for an
arbitrary microscopic free energy. This includes the case where $U_{\Lambda}$
has several distinct minima and, in particular, the interesting case of a
tricritical point. At present, the main inaccuracy arises from a
simplification of the $q^2$-dependence of the four point function which
reflects itself in an error in the anomalous dimension $\eta$.
The simplification of the momentum dependence of the effective propagator
in the flow equation plays presumably only a secondary role.
In summary, the non-perturbative flow equation appears to be a very
efficient tool for the establishment of an explicit quantitative
connection between the microphysical interactions and the
long-range properties of the free energy.

\begin{flushleft}
Acknowledgment:  We would like to thank B. Bergerhoff, J. Berges and
M. Tsypin for helpful discussions.
\end{flushleft}

% ********************************************************************

\end{document}